\documentclass[preprint,aps,12pt,nofootinbib,tightenlines,amsmath,amssymb]{revtex4}

\usepackage{amsmath}
\usepackage{graphicx}
\usepackage{amssymb}
\usepackage{color}

\newcommand{\be}{\begin{eqnarray}}
\newcommand{\ee}{\end{eqnarray}}

\newcommand{\wh}{\widehat}
\newcommand{\p}{\partial}

\newcommand{\dbar}{\lower0.1ex\hbox{$\mathchar'26$}\mkern-12mu d}
 \newcommand{\kbar}{\lower0.1ex\hbox{$\mathchar'26$}\mkern-10mu k}

\newcommand{\nn}{\nonumber}

\newcommand{\Tr}{\mathop{\rm Tr}\nolimits}
\newcommand{\diag}{\mathop{\rm diag}}
\newcommand{\cL}{{\mathcal L}}

\newcommand{\cA}{{\mathcal A}}
\newcommand{\hcA}{ \hat{{\mathcal A}}}

\newcommand{\hcF}{ \hat{{\mathcal F}}}

\newcommand{\bs}{\boldsymbol}

\newcommand{\bscF}{ \boldsymbol{{\mathcal F}}}
\newcommand{\bscA}{\boldsymbol{{\mathcal A}}}

\newcommand{\hcW}{ \hat{{\mathcal W}}}

\newcommand{\bscM}{ \boldsymbol{{\mathcal M}}}

\newcommand{\hA}{\hat{A}}

\newcommand{\bshcF}{ \boldsymbol{\hat{\mathcal F}}}
\newcommand{\bshcA}{\boldsymbol{\hat{\mathcal A}}}
\newcommand{\bshcD}{ \boldsymbol{\hat{\mathcal D}}}
\newcommand{\bshcW}{ \boldsymbol{\hat{\mathcal W}}}

\newcommand{\hF}{\hat{F}}

\newcommand{\cF}{{\mathcal F}}

\newcommand{\sF}{{\mathsf F}}
\newcommand{\hsF}{{\hat{\sF}}}

\newcommand{\whsF}{{\widehat{\sF}}}

\newcommand{\mV}{{\mathsf V}}
\newcommand{\cV}{{\mathcal V}}

\newcommand{\cD}{{\mathcal D}}

\newcommand{\hcD}{\hat{{\mathcal D}}}

\newcommand{\cG}{{\mathcal G}}

\newcommand{\whcG}{\widehat{{\mathcal G}}}

\newcommand{\mJ}{\mathrm J}

\newcommand{\whmJ}{\widehat{\mJ}}

\newcommand{\cW}{{\mathcal W}}
\newcommand{\cS}{{\mathcal S}}

\newcommand{\bchi}{\bar{\chi} }

\newcommand{\1}{\mspace{1mu}}

\newcommand{\tga}{\tilde{\gamma}}

\newcommand{\cR}{\mathcal R}

\newcommand{\mT}{\mathsf T}

\newcommand{\mM}{\mathsf M}

\newcommand{\adjoin}{\,\widetilde{\Join}\, }

\newcommand{\mC}{\mathsf C}

\newcommand{\bM}{\bar{M}}

\newcommand{\mOm}{\mathsf \Omega }

\newcommand{\hmOm}{\hat{\mOm}}

\newcommand{\chih}{\hat{\chi}}

\newcommand{\tchi}{\tilde{\chi}}

\newcommand{\etat}{\tilde{\eta}}
\newcommand{\etab}{\bar{\eta}}

\newcommand{\etach}{\check{\eta}}

\newcommand{\mG}{\mathsf{G}}
\newcommand{\cP}{\mathcal{P}}
\newcommand{\cC}{\mathcal{C}}

\newcommand{\cT}{\mathcal{T}}
\newcommand{\hcT}{\hat{\cT}}

\newcommand{\chib}{\bar{\chi}}

\begin{document}
\def\intdk{\int\frac{d^4k}{(2\pi)^4}}
\def\sla{\hspace{-0.22cm}\slash}
\hfill

%\begin{titlepage}

\title{Gravitization Equation and Zero Energy Momentum Tensor Theorem with Cancellation Law in Gravitational Quantum Field Theory}

\author{Yue-Liang Wu}\email{ylwu@itp.ac.cn, ylwu@ucas.ac.cn}
\affiliation{$^1$ Institute of Theoretical Physics, Chinese Academy of Sciences, Beijing 100190, China\\
$^2$ International Centre for Theoretical Physics Asia-Pacific (ICTP-AP), UCAS, Beijing 100190, China \\
$^3$ Taiji Laboratory for Gravitational Wave Universe (Beijing/Hangzhou), University of Chinese Academy of Sciences (UCAS), Beijing 100049, China \\
$^4$ School of Fundamental Physics and Mathematical Sciences, Hangzhou Institute for Advanced Study, UCAS, Hangzhou 310024, China }

%\date{\today}

\begin{abstract}
We investigate the essential properties of gravitational quantum field theory (GQFT) based on spin gauge symmetry, using the general theory of quantum electrodynamics as an example. A constraint equation for the field strength of the gravigauge field is derived, serving as a gravitization equation within the spin-related gravigauge spacetime. This equation reveals how gravitational effects emerge from the non-commutative relation of the gravigauge derivative operator. By transmuting the action from gravigauge spacetime to Minkowski spacetime, we demonstrate that translational invariance results in a vanishing energy-momentum tensor in GQFT when the equations of motion are applied to all fundamental fields, including the gravigauge field. This extends the conservation law of the energy-momentum tensor in quantum field theory to a cancellation law of the energy-momentum tensor in GQFT. As a result, an equivalence between the general gravitational equation and the zero energy-momentum tensor theorem naturally arises in GQFT. Certain aspects of the Poincar\'e gauge theory are also briefly discussed. Furthermore, a GQFT incorporating the Chern-Simons action in three-dimensional spacetime is developed, based on the inhomogeneous spin gauge symmetry WS(1,2) and the global Poincar\'e symmetry PO(1,2). This framework provides a basis for exploring its connection to Witten's perspective on three-dimensional gravity.
﻿\end{abstract}

%\pacs{04.50.+h,11.15.-q,12.20.-m} 

\maketitle

\begin{widetext}
%\tableofcontents
\end{widetext}

\section{Introduction}

The framework of quantum field theory (QFT) was initially developed to describe quantum electrodynamics (QED) \cite{QED1,QED2a,QED2b,QED3a,QED3b,QED3c,QED4a,QED4b}, successfully combining quantum mechanics and special relativity. It serves as a theoretical foundation for describing the microscopic world. QFT underpins our understanding of fundamental forces and elementary particles. In this framework, elementary particles, as the basic constituents of matter, are represented as quantum fields with specific intrinsic properties. This concept replaces Newton's hypothesis of point-like particles, which treated matter as dimensionless points without internal structure. The intrinsic properties of quantum fields are characterized by quantum numbers, such as spin, electric charge, isospin, and color-spin. These quantum numbers determine the gauge symmetries that govern the fundamental interactions of elementary particles, in accordance with the principle of gauge invariance.

QFT has successfully described the three fundamental forces, electroweak and strong interactions, within the Standard Model (SM) \cite{EW1,EW2,EW3,QCD1,QCD2,TW}, based on gauge field theories \cite{GT1,GT2}. To date, QFT has proven highly effective in addressing the microscopic world of elementary particles, including their behavior and fundamental interactions at the quantum level.

Einstein's general relativity (GR) \cite{GR,FGR} was established to describe gravity. In this theory, gravity is no longer treated as an instantaneous long-range force but rather as the dynamics of curved spacetime caused by mass and energy. This revolutionary framework redefined our understanding of spacetime and gravitation by abandoning Newton's hypothesis of long-range instantaneous interactions. GR has also proven highly effective in describing the macroscopic world of gravitation and the behavior of massive objects within the framework of spacetime curvature on large scales.

Nevertheless, merging GR and QFT presents a profound challenge, as it requires reconciling their fundamentally different perspectives. To harmonize the contrasting notions inherent in these two theoretical frameworks, a quantum field theory of gravity was developed based on the spin gauge symmetry SP(1,3). This provides a theoretical foundation referred to as Gravitational Quantum Field Theory (GQFT) \cite{GQFT1,GQFT2}, which treats gravitational interactions on the same footing as electroweak and strong interactions.

In GQFT, the key to reconciling GR and QFT lies in a fundamental principle: the laws of nature are governed by the intrinsic properties of the basic constituents of matter. This principle necessitates a rigorous distinction between intrinsic symmetries, determined by the quantum numbers of elementary particles as quantum fields, and external symmetries, which describe their motion in the flat Minkowski spacetime of coordinates. This crucial distinction motivates the conceptualization of a bi-frame spacetime with a fiber bundle structure. Here, the globally flat Minkowski spacetime serves as the base spacetime, describing the free motion of elementary particles as quantum fields, while the spin-related intrinsic gravigauge spacetime acts as the fiber, representing an observable interacting spacetime that characterizes fundamental interactions among elementary particles.

This conceptual framework highlights the distinguishable features between external and intrinsic symmetries, leading to a comprehensive understanding of the dynamics governing the fundamental interactions of elementary particles as quantum fields. As a consequence, the global Lorentz symmetry SO(1,3) in Minkowski spacetime and the intrinsic spin symmetry SP(1,3) in the Hilbert space of the Dirac spinor field are unified as joint symmetries SO(1,3)$\Join$SP(1,3) in GQFT, rather than being treated as associated symmetries SO(1,3)$\adjoin$SP(1,3) in QFT. Here, the spin symmetry SP(1,3) is localized to be an intrinsic gauge symmetry, along with the gauge invariance principle. To ensure the joint symmetries SO(1,3)$\Join$SP(1,3), a spin-related gravigauge field $\chi_{\mu}^{\; \, a}(x)$, as a bi-covariant vector field defined in bi-frame spacetime, is introduced as the basic gravitational field, replacing the metric field in GR, to describe fundamental gravitational interactions. Since the GQFT framework provides a unified description of gravitational, electroweak, and strong interactions, it enables the construction of a hyperunified field theory aimed at unifying all fundamental interactions and elementary particles (see refs.\cite{HUFT1,HUFT2,FHUFT1,FHUFT2,FHUFT} and references therein).
 
 As the spin-related gravigauge field behaves as a Goldstone-type boson, it is identified as a massless graviton, and its equation of motion is used to describe gravidynamics within the framework of GQFT. Recently, it was further demonstrated that, in the absence of source fields, the linearized gravidynamics of the gravigauge field leads to an intriguing prediction regarding gravitational waves. Specifically, GQFT predicts five independent polarizations of gravitational waves\cite{GQFT3}, in contrast to the two polarizations of the spin-2 mode in GR. The additional three polarizations in GQFT arise from a spin-0 mode with one polarization and a spin-1 mode with two polarizations. The presence of the spin-1 mode in GQFT implies a breakdown of the strong equivalence principle postulated in GR.
 
In this paper, we aim to further explore some essential properties within the framework of GQFT. A gravitization equation emerges from the constraint equation for the field strength of the gravigauge field in spin-related gravigauge spacetime. This field strength of the gravigauge field reflects the group structure factor of the non-commutative gravigauge derivative operator defined in gravigauge spacetime. Notably, the gravigauge field acts as an auxiliary field in the action of the general theory of QED formulated in local orthogonal gravigauge spacetime. Interestingly, we demonstrate that translational invariance in Minkowski spacetime leads to a cancellation law for the energy-momentum tensor in GQFT. This result essentially extends the conservation law of the energy-momentum tensor in conventional QFT. As a consequence, we derive a general theorem stating that the energy-momentum tensor vanishes throughout the entire Minkowski spacetime in the presence of gravitational interaction mediated by the gravigauge field in GQFT. This theorem may be termed the zero energy-momentum tensor theorem.

We would like to address that the bi-covariant vector field $\chi_{\mu}^{\; \, a}(x)$ dual to the invertible bi-covariant vector field $\chih_{a}^{\; \mu}$ introduced in association with the spin gauge symmetry SP(1,3) of the spinor field manifests as a gauge-type vector field in the description of gravitational interaction through its field strength in GQFT. This approach differs conceptually from most gravity gauge theories proposed in earlier studies \cite{GGT1,GGT2,GGT3,GGT4,GGT5}, which have been extensively explored over the past half-century. Detailed descriptions and critiques of these theories can be found in review articles \cite{GGTR1,GGTR2,GGTR3,GGTR4} and references therein. It is noteworthy that most gravity gauge theories were formulated based on Riemannian or non-Riemannian geometry on curved spacetime, or on the Poincar\'e gauge group/conformal gauge group, with the introduction of a vector bundle isomorphic to the tangent bundle in the spacetime manifold of coordinates. Notably, since GR becomes trivial in three-dimensional spacetime, a three-dimensional gravity gauge theory was proposed based on the inhomogeneous Lorentz gauge group symmetry ISO(1,2) \cite{GGT3D1}, where a purely Chern-Simons action was anticipated to be exactly solvable at both classical and quantum levels when disentangling the Hamiltonian constraint equations.

However, in all these cases, the local group symmetries are treated as external symmetries rather than internal symmetries arising from the intrinsic quantum numbers of the fundamental constituents of matter. Consequently, the fundamental questions regarding the definition of space and time, as well as the quantization of gravity gauge theories, remain unresolved. Additionally, the basic action structure, dynamic properties, and interactions of gravity gauge theories with fundamental spinor fields are not yet fully understood. For clarity, we will provide more detailed analyses and discussions on these issues later in the text.

\section{General Theory of QED and Gravitization Equation in GQFT}

For the purpose of demonstration on intriguing properties within the framework of GQFT, let us extend by example the QED described by QFT to a general theory of QED within the framework of GQFT. 

For simplicity, we begin with the following action of QED in framework of QFT:
\be \label{actionQED}
S_{QED} = \int d^4x\,  \frac{1}{2} \left( \bar{\psi}(x) \gamma^{a} \delta_{a}^{\;\;\mu} i D_{\mu} \psi(x) + H.c. \right)  - m\, \bar{\psi}(x) \psi(x)  - \frac{1}{4} \eta^{\mu\mu'}\eta^{\nu\nu'} F_{\mu\nu}F_{\mu'\nu'}, 
\ee
where $\psi(x)$ represents complex Dirac spinor field\cite{DE} (for instance, electron) with mass $m$. $\gamma^{a}$ (a=0,1,2,3) are Dirac $\gamma$-matrices and $\delta_{a}^{\;\;\mu} $ is the Kronecker symbol. The Greek alphabet ($\mu,\nu = 0,1,2,3$) and Latin alphabet ($a,b,=0,1,2,3$) are introduced to distinguish coordinate vectors and spin vectors, where the Greek and Latin indices are raised and lowered by the constant metric tensors $\eta^{\mu\nu}$($\eta_{\mu\nu}$) =$\diag.$(1,-1,-1,-1) and $\eta^{ab}$($\eta_{ab}$)=$\diag.$(1,-1,-1,-1). The covariant derivative $D_{\mu}$ and field strength $F_{\mu\nu}$ are the usual ones:
\be \label{CD}
 & & iD_{\mu} \equiv  i\p_{\mu} + g_e A_{\mu}(x) , \nn \\
 & & F_{\mu\nu} = \p_{\mu}A_{\nu} - \p_{\nu} A_{\mu} , 
\ee
where $A_{\mu}(x)$ is the electromagnetic gauge field with coupling constant $g_e$.

The above action is known to have an associated symmetry:
\be
G_S = PO(1,3)\adjoin SP(1,3) = P^{1,3}\ltimes SO(1,3)\adjoin SP(1,3) ,
\ee
where PO(1,3) $\equiv P^{1,3}\ltimes$SO(1,3) denotes Poincar\'e symmetry group (or inhomogeneous Lorentz symmetry) with P$^{1,3}$ the translational symmetry in Minkowski spacetime. The symbol ``$\adjoin$" is used to notate the associated symmetry, namely, the transformation of spin symmetry SP(1,3) in Hilbert space of Dirac fermion must be coincidental to that of the isomorphic Lorentz symmetry SO(1,3) in Minkowski spacetime of coordinates. Actually,  it is well known that such an associated symmetry lays the foundation in deriving Dirac equation of electron.  

The essential notion in GQFT is to distinguish the group symmetries in internal Hilbert space and external Minkowski spacetime by following along the gauge invariance principle. It is stated via the gauge invariance principle that {\it the laws of nature should be independent of the choice of local field configurations}. Namely, all internal symmetries should be localized as gauge symmetries to characterize fundamental interactions. Grounded on the gauge invariance principle, the internal spin symmetry SP(1,3) for Dirac fermion $\psi(x)$ must be a local gauge symmetry. While the Poincar\'e group symmetry PO(1,3) of coordinates is considered to remain a global symmetry. 

In general, a spin gauge field $\cA_{\mu}(x)$ is introduced to preserve spin gauge symmetry SP(1,3). Unlike usual internal symmetry, to keep both local spin gauge symmetry SP(1,3) and global Lorentz symmetry SO(1,3), an additional bi-covariant vector field $\chih_{a}^{\; \mu}(x)$ is required to replace the Kronecker delta symbol $\delta_{a}^{\;\;\mu}$. Consequently, the ordinary covariant derivative $D_{\mu}$ associated with the delta symbol $\delta_{a}^{\;\;\mu}$ in Eq.(\ref{CD}) should be extended to a general covariant derivative as follows:  
\be
& & i \delta_{a}^{\;\;\mu} D_{\mu}  \to i\chih_{a}^{\; \, \mu} \cD_{\mu} \equiv i\hcD_{a} 
\ee
where we have adopted the following definition:
\be
 i\hcD_{a} & \equiv & \chih_{a}^{\; \, \mu}( iD_{\mu} + g_s\cA_{\mu} ) \nn \\
& \equiv & i\eth_{a} + g_e \hA_{a} + g_s\hcA_{a}, 
\ee
with
\be
\cA_{\mu} & \equiv& \cA_{\mu}^{ab} \frac{1}{2} \Sigma_{ab} ,\quad \Sigma_{ab} = \frac{i}{4}[\gamma_a, \gamma_b], \nn \\
 \hcA_{a} & \equiv & \chih_{a}^{\; \, \mu} \cA_{\mu}, \quad \hA_{a} \equiv \chih_{a}^{\; \, \mu} A_{\mu},
\ee
where $\Sigma_{ab}$ represents the group generators of spin gauge symmetry SP(1,3). It is noticed that $\eth_{a}$ is regarded as a spin-related intrinsic derivative operator, which motivates us to introduce a corresponding displacement vector $\dbar\zeta^{a}$. They are related to the ordinary coordinate derivative operator $\partial_{\mu}\equiv \frac{\partial}{\partial x^{\mu}}$ and displacement vector $dx^{\mu}$ via the dual bi-covariant vector fields $\chih_{a}^{\; \mu}(x)$ and $\chi_{\mu}^{\; a}(x)$ as follows:
\be
 \eth_{a}  \equiv  \chih_{a}^{\; \mu}(x)\p_{\mu}, \quad \dbar\zeta^{a} \equiv  \chi_{\mu}^{\; a}(x) dx^{\mu}, 
\ee
with $\chih_{a}^{\; \mu}(x)$ and $\chi_{\mu}^{\; a}(x)$ satisfying the dual conditions:
\be
\chi_{\mu}^{\; a}(x)  \eta_{a}^{\; b} \chih_{b}^{\; \nu}(x) = \eta_{\mu}^{\; \nu} , \quad 
\chih_{b}^{\; \nu}(x) \eta_{\nu}^{\; \mu} \chi_{\mu}^{\; a}(x)  = \eta_{b}^{\; a} .
\ee
Where $\chih_{a}^{\; \mu}(x)$ and $\chi_{\mu}^{\; a}(x)$ transform as bi-covariant vector fields under transformations of the spin gauge symmetry SP(1,3) and global Lorentz symmetry SO(1,3):
\be \label{GT}
& & \chi_{\mu}^{\; a}(x) \to \chi_{\mu}^{'\; a}(x) =  \Lambda^{a}_{\; b}(x)\chi_{\mu}^{\; b}(x), \;\; \chih_{a}^{\; \mu}(x) \to \chih_{a}^{'\; \mu}(x) =  \Lambda_{a}^{\; b}(x) \chih_{b}^{\; \mu}(x),
\ee
and
\be
& & \chi_{\mu}^{\; a}(x) \to \chi_{\mu}^{'\; a}(x') =  L_{\mu}^{\; \nu}\, \chi_{\nu}^{\; a}(x), \;\; \chih_{a}^{\; \mu}(x) \to \chih_{a}^{'\; \mu}(x') =  L^{\mu}_{\; \nu}\, \chih_{a}^{\; \nu}(x) ,\nn \\
& & x^{' \mu} = L^{\mu}_{\; \nu}\, x^{\nu} , \quad L^{\mu}_{\; \nu} \in SO(1,3), \;\;  \Lambda^{a}_{\; b}(x) \in SP(1,3) \cong SO(1,3) .
\ee

It is known that the derivative operator $\p_{\mu}$ and displacement vector $dx^{\mu}$ constitute dual bases $\{\p_{\mu}\}\equiv \{\partial/\partial x^{\mu}\}$ and $\{dx^{\mu}\}$, which span tangent spacetime $\mT_{\mM}$ and cotangent spacetime $\mT^{*}_{\mM}$, respectively, and mentioned as external spacetime of coordinates. In analogues, the intrinsic derivative operator $\eth_a$ and displacement vector $\dbar \zeta^{a}$ define dual bases $\{\eth_a\}$ and $\{\dbar \zeta^{a}\}$, giving rise to spin-related intrinsic spacetime $\mT_{\mG}$ and dual spacetime $\mT^{*}_{\mG}$. 

The spin-related intrinsic basis $\{\eth_a \}$ does not commute and satisfies the following commutation relation:
\be \label{NCR}
& & [ \eth_c ,\; \eth_d] = \hsF_{cd}^a \eth_a ,
\ee
with $\hsF_{cd}^a$ regarded as group structure factor which gets the following explicit form:
\be \label{LGSF}
& &  \hsF_{cd}^a \equiv  - \chih_{c}^{\; \mu} \chih_{d}^{\; \nu} \sF_{\mu\nu}^{a} , \quad \sF_{\mu\nu}^{a} \equiv \p_{\mu}\chi_{\nu}^{\; a}(x) - \p_{\nu}\chi_{\mu}^{\; a}(x) ,
\ee 
which indicates that the group structure factor $\hsF_{cd}^a$ is characterized by a gauge-type field strength, $\sF_{\mu\nu}^{a}$, which is defined via a gauge-type bi-covariant vector field $\chi_{\mu}^{\; a}(x)$. 

From the above analyses, the spin-related gauge-type bi-covariant vector field $\chi_{\mu}^{\; a}(x)$ is inevitably introduced to describe the gravitational interaction when a spinor field is taken as basic constituent of matter. Therefore, the bi-covariant vector field $\chi_{\mu}^{\; a}(x)$ is designated as {\it gravigauge field } and its field strength $\sF_{\mu\nu}^{a}$ is considered as {\it gravigauge field strength}. In general, $\chi_{\mu}^{\; a}(x)$ is regarded as a gauge field defined in Minkowski spacetime and valued in spin-related intrinsic spacetime, where the global flat Minkowski spacetime is taken as a base spacetime and the spin-related intrinsic spacetime is viewed as a fiber. Such an intrinsic spacetime characterized by gravigauge field is referred to as {\it gravigauge spacetime}. In correspondence, the spin-related intrinsic derivative operator $\eth_{a}$ and displacement vector $\dbar \zeta^{a}$ are called as {\it gravigauge derivative} and {\it gravigauge displacement}. 

The intrinsic gravigauge spacetime spanned by spin-related gravigauge bases $\{\eth_a\}$ and $\{\dbar \zeta^{a}\}$ facilitates a non-commutative geometry of spin fiber, which distinguishes from the external flat Minkowski spacetime characterized by commutative geometry of coordinates. It is not difficult to show that the spin-related gravigauge spacetime can also be characterized by a spin connection $\mOm_{\mu}^{ab}$ determined by the gravigauge field as follows: 
\be \label{SGGF}
& & \mOm_{\mu}^{ab}(x) = \frac{1}{2}\left( \chih^{a\nu} \sF_{\mu\nu}^{b} - \chih^{b\nu} \sF_{\mu\nu}^{a} -  \chih^{a\rho}  \chih^{b\sigma}  \sF_{\rho\sigma}^{c} \chi_{\mu c } \right) 
\ee
which may be referred to as {\it spin gravigauge field}. The spin gravigauge field $\mOm_{\mu}^{ab}$ has the same transformation property as the spin gauge field $g_s\cA_{\mu}^{ab}$ under transformations of spin gauge symmetry SP(1,3). The group structure factor $\hsF_{cd}^a$ and spin gravigauge field $\mOm_{\mu}^{ab}$ get the following relations:
\be
& & \hsF_{cd}^{a}  = \hmOm_{cd}^{a}  - \hmOm_{dc}^{a} \equiv \hmOm_{[cd]}^{a}, \quad \hmOm_{c}^{ab} \equiv \chih_{c}^{\; \mu} \mOm_{\mu}^{ab} = \etach_{ca'}^{[ab] c'd'} \hsF_{c'd'}^{a'}  , \nn \\
& & \etach_{ca'}^{[ab] c'd'} \equiv \frac{1}{2}  (\eta^{ac'}\eta_{a'}^{\, b} - \eta^{bc'}\eta_{a'}^{\, a}) \eta_{c}^{\, d'} + \frac{1}{4} (\eta^{ac'}\eta^{bd'} - \eta^{bc'}\eta^{ad'} ) \eta_{ca'} .
\ee

In terms of the general covariant derivative $\bshcD_{c}=\eth_c + g \bshcA_{c}$, it enables us to define the gauge covariant field strength $\bshcF_{cd}$ via the commutation relation, 
\be
[ i\bshcD_c ,\; i\bshcD_d] = i  g ( \bshcF_{cd} + \hsF_{cd}^a\, i\bshcD_a ) ,
\ee
with $\bshcF_{cd}$ given by the following general form:
\be
& & \bshcF_{cd} \equiv  \cD_{c}\bshcA_{d} - \cD_{d}\bshcA_{c} - i g [\bshcA_{c}, \bshcA_{d}] \equiv \bscF_{cd} + \hsF_{cd}^{a} \bshcA_{a} , \nn \\
& & \cD_{c}\bshcA_{d} \equiv \eth_{c} \bshcA_{d} + \hmOm_{cd}^{a} \bshcA_{a} , 
  \; \; \hsF_{cd}^{a} \bshcA_{a} \equiv (\hmOm_{cd}^{a} - \hmOm_{dc}^{a}) \bshcA_{a}, \nn \\
& & \bscF_{cd} \equiv \eth_{c}\bshcA_{d} - \eth_{d}\bshcA_{c} - i g [\bshcA_{c}, \bshcA_{d}] , \quad \bshcF_{cd} \equiv \chih_{c}^{\; \mu} \chih_{d}^{\; \nu} \bscF_{\mu\nu}  ,
\ee
where the extra term $\hsF_{cd}^{a}\bshcA_{a}$ reflects gravitational effect due to non-commutative feature in spin-related gravigauge spacetime. In our present considerations, it involves two gauge field strengths, $\bshcF_{cd} = \hcF_{cd}$ ($\bshcA_{c}=\hcA_{c}$, $g=g_s$) and $\bshcF_{cd} =\hF_{cd}$ ($\bshcA_{c}=\hA_{c}$, $g=g_e$), which correspond to the spin gauge field strength and electromagnetic gauge field strength, respectively. 

The introduction of gravigauge field, $\chi_{\mu}^{\; a}(x)$, becomes a necessity to construct a spin gauge invariant action when utilizing the gravigauge derivative operator, $\eth_{a}$, and displacement vector, $\dbar \zeta^{a}$, in spin-related gravigauge spacetime. Grounded on the internal gauge symmetries:
\be
\mG_{S} =  U(1)\times SP(1,3)
\ee
we are able to build a general theory of QED in spin-related gravigauge spacetime as follows:
\be \label{actionGQED}
\cS_{GQED}  & = & \int [\dbar \zeta^{c}]\,  \frac{1}{2} \left( \bar{\psi}(x) \gamma^{c} i \hcD_{c} \psi(x) + H.c. \right)  - m\, \bar{\psi}(x) \psi(x)  \nn \\
& - & \frac{1}{4} \eta^{c c'}\eta^{d d'} ( \hF_{cd}\hF_{c'd'} + 2 \Tr \hcF_{cd}\hcF_{c'd'} )  \nn \\
& + &  \frac{1}{4} \bM_{\kappa}^2 \etat^{cdc'd'}_{aa'} \hsF_{cd}^{a}\hsF_{c'd'}^{a'}  +   M_{\cA}^2 \eta^{cd}  \Tr (\hcA_{c} -  \hmOm_{c}/g_s)( \hcA_{d} -  \hmOm_{d}/g_s),
\ee
where $\bM_{\kappa}$ represents fundamental mass scale and $M_{\cA}$ denotes the mass of spin gauge field. The constant tensor $\etat^{cdc'd'}_{aa'}$ has a special structure in order to ensure the spin gauge symmetry SP(1,3) for the quadratic form of gravigauge field strength $\hsF_{cd}^{a}$, its explicit form is given as follows:
\be \label{STensor1}
\tilde{\eta}^{cd c'd'}_{a a'}  & \equiv &    \eta^{c c'} \eta^{d d'} \eta_{a a'}  
+  \eta^{c c'} ( \eta_{a'}^{d} \eta_{a}^{d'}  -  2\eta_{a}^{d} \eta_{a'}^{d'}  ) +  \eta^{d d'} ( \eta_{a'}^{c} \eta_{a}^{c'} -2 \eta_{a}^{c} \eta_{a'}^{c'} ) . 
\ee

It is interesting to observe that there exists no dynamic term for the gravigauge field strength, $\hsF_{cd}^a$, in the above action Eq.(\ref{actionGQED}), which indicates that $\hsF_{cd}^a$ emerges as an auxiliary field. This observation enables to obtain, from the least action principle, the following relation:
\be \label{GE}
\bscM_{cda}^{\; c'd'a'} \hsF_{c'd' a'}  = \widehat{\bscF}_{cda} ,  \quad \mbox{or}\quad \hsF_{cd a}  = \widehat{\bscM}_{cda}^{\; c'd'a'} \widehat{\bscF}_{c'd'a'}
\ee
which provides a constraint equation for the gravigauge field strength (or group structure factor) $\hsF_{cd}^a$ in gravigauge spacetime. The explicit expressions for $\bscM_{cda}^{\; c'd'a'}$ and $\widehat{\bscF}_{cda}$ are found to have the following forms:
\be \label{GEM}
\bscM_{cda}^{\; c'd'a'}  &\equiv & \etab_{cda}^{\; c'd'a'} \bM^2_{\kappa} +  \eta_{cda}^{\; c'd'a'} M_{-}^2 + \eta_{cd}^{\; c'd'} ( \eta_{a}^{\, a'}  M_{+}^{2}   - \widehat{\cV}_{a b} \eta^{b a'} ) , \nn \\
\widehat{\bscF}_{cda}  & \equiv &  \cF_{cd}^{a'b'}\hcA_{a a'b'} +  F_{cd} \hA_{a} - g_s^{-1}  M^2_{\cA} [ \hcA_{acd}  + 2(\hcA_{cda}-\hcA_{dca} ) ]  ,
\ee
with
\be
& &  \etab_{cda}^{\; c'd'a'} \equiv ( \eta_{c}^{\, d'}  \eta^{c'a'} -\eta_{c}^{\, c'}  \eta^{d'a'} ) \eta_{da} - (\eta_{d}^{\, d'}  \eta^{c'a'} - \eta_{d}^{\, c'}  \eta^{d'a'}) \eta_{ca} , \nn \\
& & \eta_{cda}^{\; c'd'a'} \equiv \frac{1}{2} [ ( \eta_{c}^{\, c'}  \eta_{d}^{\, a'}  - \eta_{d}^{\, c'} \eta_{c}^{\, a'} ) \eta_{a}^{\, d'} +  (\eta_{d}^{\, d'} \eta_{c}^{\, a'} - \eta_{c}^{\, d'} \eta_{d}^{\, a'} )\eta_{a}^{\, c'} ] , \nn \\
& &  \eta_{cd}^{\; c'd'} \equiv \frac{1}{2} (\eta_{c}^{\, c'} \eta_{d}^{\, d'} -\eta_{d}^{\, c'} \eta_{c}^{\, d'} ) , \nn \\
& & M_{\pm}^2  \equiv \bM^2_{\kappa} +( \frac{1}{2}  \pm 1) M_{\cA}^2/g_s^2 , \;\; \widehat{\cV}_{a b}  \equiv  \hcA_{a a'b'}\hcA_{b}^{a'b'} + \hA_{a}\hA_{b},
\ee
where $\bscM_{cda}^{\; c'd'a'}$ is viewed as a $24\times 24$ matrix, antisymmetric under exchanges of $c, d$ and $c', d'$. $\widehat{\bscM}_{cda}^{\; c'd'a'} $ denotes its inverse matrix. 

The constraint equation shows that the gravigauge field strength $\hsF_{cd}^a$ is intricately governed by the collective dynamics of spin gauge field and electromagnetic gauge field, so we may designate such an equation as {\it gravitization equation}. This may be understood intuitively as that the gravigauge spacetime as a fiber is considered as a local orthogonal spacetime characterized by the spin-related gravigauge field. In such a local orthogonal gravigauge spacetime, the gravitational effect emerges as non-commutative nature of gravigauge derivative operator, and also the gravitational interaction revealed from the group structure factor $\hsF_{cd}^a$ appears as an auxiliary field.

\section{Zero Energy Momentum Tensor Theorem with Cancelation Law in GQFT}

The dual gravigauge fields $\chi_{\mu}^{\;\; a}$ and $\chih_{a}^{\;\, \mu}$ as bi-covariant vector fields in bi-frame spacetime behave as Goldstone bosons, which enables to represent the above action Eq.(\ref{actionGQED}) into the following form within the framework of GQFT:
\be \label{actionGQED2}
\cS_{GQED}  & = & \int [d x]\chi \cL_{GQED} \nn \\
& = & \int d^4x\,  \chi \{ \frac{1}{2} \left( \bar{\psi}(x) \gamma^{a} \chih_{a}^{\; \mu} i\cD_{\mu} \psi(x) + H.c. \right)  - m\, \bar{\psi}(x) \psi(x)  \nn \\
& - & \frac{1}{4} \chih^{\mu\mu'}\chih^{\nu \nu'} ( F_{\mu\nu}F_{\mu'\nu'} + \cF_{\mu\nu}^{ab}\cF_{\mu'\nu' ab} ) + \frac{1}{4} \bM_{\kappa}^2 \tchi^{\mu\nu\mu'\nu'}_{aa'} \sF_{\mu\nu}^{a}\sF_{\mu'\nu'}^{a'}  \nn \\
& + &   \frac{1}{2} M_{\cA}^2 \chih^{\mu\nu} (\cA_{\mu}^{ab} -  \mOm_{\mu}^{ab}/g_s)( \cA_{\nu ab} -  \mOm_{\nu ab}/g_s) \} ,
\ee
with the tensors defined as: 
\be
& & \tchi_{aa'}^{\mu\nu \mu'\nu'} \equiv \chih_{c}^{\;\, \mu}\chih_{d}^{\;\, \nu} \chih_{c'}^{\;\, \mu'} \chih_{d'}^{\;\, \nu'}  \etat^{c d c' d'}_{a a'} , \nn \\
& &  \chih^{\mu\nu} \equiv \chih_{a}^{\; \mu} \chih_{b}^{\; \nu} \eta^{ab} , \quad \chi_{\mu\nu} \equiv \chi^{\;a}_{\mu} \chi^{\; b}_{\nu} \eta_{ab} , \nn \\
& & \chi = \det \chi_{\mu}^{\; a} = \sqrt{-\det \chi_{\mu\nu}}, 
\ee  
where the tensor $\chi_{\mu\nu}$ is viewed as a symmetric composed field and referred to as {\it gravimetric field}.

By applying the least action principle to various fields, we are able to derive their equations of motion. The equation of motion for the Dirac fermion $\psi(x)$ is obtained to be,
\be \label{EoMD}
\gamma^{a} \chih_{a}^{\; \mu} i( \cD_{\mu}  - \mV_{\mu}) \psi(x) - m \psi(x) = 0, 
\ee
with $\mV_{\nu}$ regarded as induced vector gauge field,
\be
\mV_{\mu} & \equiv & \frac{1}{2} \chi \, \chih_{b}^{\;\; \nu}\cD_{\nu}(\chih \chi_{\mu}^{\;\; b}), \;\; \cD_{\nu}(\chih\chi_{\mu}^{\;\;a}) = \p_{\nu} (\chih\chi_{\mu}^{\;\; a}) + \chih g_s\cA_{\nu\, b}^{a}  \chi_{\mu}^{\;\;b},
\ee
which reflects a novel effect caused from gravigauge field and spin gauge field. 

For the electromagnetic gauge field $A_{\mu}$, its equation of motion is given by:
\be \label{EoMEM}
& & \p_{\nu} \widehat{F}^{\mu\nu} = -\wh{J}^{\mu} ,
\ee
with the field strength and conserved current defined as follows:
\be \label{GFS}
& &  \widehat{F}^{\mu\nu} \equiv \chi \hat{\chi}^{[\mu\nu]\mu'\nu'}  F_{\mu'\nu'} , \;\; \hat{\chi}^{[\mu\nu]\mu'\nu'} \equiv \frac{1}{2} ( \chih^{\mu\mu'}\chih^{\nu\nu'} - \chih^{\nu\mu'}\chih^{\mu\nu'} ) , \nn \\
& & \p_{\mu} \wh{J}^{\mu} = 0, \quad \wh{J}^{\mu}  \equiv \chi \chih_{a}^{\; \mu} J^{a}   , \;\; J^{a} = - g_e \bar{\psi} \gamma^{a} \psi .
\ee

The equation of motion for the spin gauge field $\cA_{\mu}^{ab}$ is derived to have the following explicit form: 
\be \label{EoMSG}
& & \cD_{\nu} \wh{\cF}_{ab}^{\mu\nu} + M_{\cA}^2 \chi \chih^{\mu\nu} (\cA_{\nu ab} - \mOm_{\nu ab}/g_s ) = \wh{J}_{ab}^{\mu} ,
\ee
with the definitions:
\be
& & \wh{\cF}_{ab}^{\mu\nu}  \equiv  \chi \chih^{\mu\mu'}\chih^{\nu\nu'} \cF_{\mu'\nu' ab} , \;\;  \cD_{\nu} \wh{\cF}_{ab}^{\mu\nu} \equiv \p_{\nu} \wh{\cF}_{ab}^{\mu\nu} - g_s \cA_{\nu a}^{c} \wh{\cF}_{cb}^{\mu\nu}- g_s \cA_{\nu b}^{c} \wh{\cF}_{ac}^{\mu\nu}, \nn \\
& & \wh{J}_{ab}^{\mu} \equiv  \chi \chih^{c\mu} g_s\bar{\psi}\Sigma_{cab}\psi  , \quad 
\Sigma_{cab} \equiv \frac{1}{4} \{ \gamma_{c}, \Sigma_{ab} \} .
\ee

As the gravitational interaction is described by the gravigauge field $\chi_{\mu}^{\; a}$ in the framework of GQFT, it always allows to choose globally flat Minkowski spacetime as a base spacetime. This is fundamentally different from GR in which the gravity is characterized by the dynamics of curved spacetime via the composite gravimetric field $\chi_{\mu\nu}$. Therefore, GQFT enables to make a meaningful definition on energy-momentum tensor through translational invariance of coordinates in global flat Minkowski spacetime. 

Let us make translational transformation of coordinates, $x^{\mu} \to x^{'\mu} = x^{\mu} + a^{\mu}$,  for arbitrary $a^{\mu}$. According to Noether’s theorem\cite{EN} that every differentiable symmetry of the action has a corresponding conservation law. The translational invariance of the action in Eq.(\ref{actionGQED2}) leads to a general expression: 
\be 
\delta S_{GQED} = \int d^4x\, \p_{\nu} ( \wh{\cT}_{\mu}^{\;\, \nu}) a^{\mu} =0  ,
\ee
with ignoring the surface terms. Where the energy-momentum tensor $ \hcT_{\mu}^{\;\, \nu}$ is found to get the following gauge invariant form:
\be \label{EMTensor}
\wh{\cT}_{\mu}^{\; \, \nu} & = & \chi \{ \eta_{\mu}^{\; \nu}  \cL_{GSM} - \frac{1}{2} ( \bar{\psi}\gamma^{a} i\cD_{\mu} \psi   + H.c.  )  \chih_{a}^{\; \nu}  \nn \\
& + &  \eta_{\mu}^{\; \; \rho}  \chih^{\nu\sigma} \chih^{\rho'\sigma'} ( F_{\rho\rho'} F_{\sigma\sigma'} + \cF_{\rho\rho'}^{ab} \cF_{\sigma\sigma' ab})  \nn \\
& - &  [ \chi_{\mu}^{\; a} ( \p_{\rho} \whsF^{\rho\nu }_{a}    + \gamma_W \cD_{\rho}\widehat{\cG}_{a}^{\rho\nu}) +  \sF_{\mu\sigma}^{a}  \whsF^{\nu\sigma}_{a}  + \gamma_W \cG_{\mu\sigma}^{a} \widehat{\cG}^{\nu\sigma}_{a}  ]/16\pi G_{\kappa} ,
\ee
with the definitions: 
\be \label{GGFS}
& & 16\pi G_{\kappa} \equiv \frac{1}{\bM_{\kappa}^2}, \quad \gamma_W \equiv  \frac{M_{\cA}^2}{g_s^2\bM_{\kappa}^2}  , \nn \\
& & \whsF^{\mu\nu }_{a} \equiv \chi \1 \tchi^{[\mu\nu]\rho\sigma}_{a b}  \sF_{\rho\sigma }^{b}  , \;\; \tchi^{[\mu\nu]\mu'\nu'}_{a a'} \equiv \frac{1}{2}( \tchi^{\mu\nu\mu'\nu'}_{a a'} - \tchi^{\nu\mu\mu'\nu'}_{a a'} ) .
\ee

In obtaining the above gauge invariant energy-momentum tensor, we have applied for the equations of motion for the Dirac spinor field in Eq.(\ref{EoMD}) and the electromagnetic gauge field in Eq.(\ref{EoMEM}) as well as the spin gauge field in Eq.(\ref{EoMSG}). Meanwhile, we have also adopted the following relation for the spin gauge invariant mass-like term:
\be \label{MSG}
 & & \frac{1}{2} \chih^{\mu\nu}  ( \cA_{\mu}^{ab}-\mOm_{\mu}^{ab} /g_s)  ( \cA_{\nu ab}-\mOm_{\nu ab}/g_s) = \frac{1}{4 g_s^2} \bchi^{\mu\nu \mu' \nu'}\cG_{\mu\nu}^{a}\cG_{\mu'\nu' a} ,  
 \ee
with the definitions:
\be \label{GCFS}
& &  \cG_{\mu\nu}^{a} \equiv  \p_{\mu} \chi_{\nu}^{\; a}  - \p_{\nu} \chi_{\mu}^{\; a} + g_s(\cA_{\mu b}^{a} \chi_{\nu}^{\; b}  - \cA_{\nu b}^{a} \chi_{\mu}^{\; b} ),  \nn \\
& & \chib_{aa'}^{\mu\nu \mu'\nu'} \equiv \chih_{c}^{\;\, \mu}\chih_{d}^{\;\, \nu} \chih_{c'}^{\;\, \mu'} \chih_{d'}^{\;\, \nu'}  \etab^{c d c' d'}_{a a'} , \nn \\
& & \etab^{c d c' d'}_{a a'} \equiv \frac{3}{2}  \eta^{c c'} \eta^{d d'} \eta_{a a'}  
+ \frac{1}{2} ( \eta^{c c'} \eta_{a'}^{d} \eta_{a}^{d'}  +  \eta^{d d'} \eta_{a'}^{c} \eta_{a}^{c'} ) ,
\ee
where $\cG_{\mu\nu}^{a}$ defines spin gauge covariant gravigauge field strength. In Eq.(\ref{EMTensor}), we have also introduced the following definition:
\be \label{GCGFS}
\whcG_a^{\mu\nu} \equiv \chi\, \bchi^{[\mu\nu]\mu'\nu'}_{a a'}  \cG_{\mu'\nu' }^{a'}, \;\; \bchi^{[\mu\nu]\mu'\nu'}_{a a'} \equiv \frac{1}{2}( \bchi^{\mu\nu\mu'\nu'}_{a a'} - \bchi^{\nu\mu\mu'\nu'}_{a a'} ) .
\ee

The above energy-momentum tensor (\ref{EMTensor}) can be rewritten into a simple form:
\be \label{EMTensor2}
\wh{\cT}_{\mu}^{\; \, \nu} & = & (\wh{\mT}_{\mu}^{\; \, \nu} -   \wh{\mG}_{\mu}^{\; \, \nu} )/16\pi G_{\kappa} , 
\ee
with the definitions:
\be \label{GC}
\wh{\mT}_{\mu}^{\; \, \nu} & \equiv & \chi_{\mu}^{\; a}\whmJ_{a}^{\; \, \nu}, \quad \wh{\mG}_{\mu}^{\; \, \nu} \equiv \chi_{\mu}^{\; a} \p_{\rho} \whsF^{\rho\nu }_{a},  \nn \\
\whmJ_{a}^{\; \nu}  & \equiv & 16\pi G_{\kappa} \chi [  \chih_{a}^{\; \nu}  \cL_{GQED}  - \chih_{b}^{\; \nu}  \chih_{a}^{\;\rho} \frac{1}{2} ( \bar{\psi}\gamma^{b} i\cD_{\rho} \psi   + H.c.  ) \nn \\
& + &  \chih_{a}^{\; \; \rho}  \chih^{\nu\sigma} \chih^{\rho'\sigma'} ( F_{\rho\rho'}F_{\sigma\sigma'} + \cF_{\rho\rho'}^{bc} \bscF_{\sigma\sigma' bc} )   ] \nn \\
 & - &  \chih_{a}^{\; \rho}  \sF_{\rho\sigma}^{b}  \whsF^{\nu\sigma}_{b} - \gamma_{W} \cD_{\rho}\widehat{\cG}_{a}^{\rho\nu} - \gamma_{W}\chih_{a}^{\; \rho}  \cG_{\rho\sigma}^{b} \widehat{\cG}^{\nu\sigma}_{b} .
\ee
The conservation law of energy-momentum tensor leads to the following relation:
\be
 \p_{\nu} \wh{\cT}_{\mu}^{\; \, \nu} = 0, \quad \p_{\nu} \whmJ_{a}^{\; \, \nu}  = (\chih_{a}^{\; \rho} \p_{\nu} \chi_{\rho}^{\; b} ) (\p_{\mu} \whsF^{\mu\nu }_{b} - \whmJ_{b}^{\; \, \nu} ) .
\ee

In the general action of QED represented in Minkowski spacetime (Eq.(\ref{actionGQED2})), the gravigauge field $\chi_{\mu}^{\;\; a}(x)$ appears as a dynamical field. By applying for the least action principle, we are able to derive the following gauge-type gravitational equation of motion with conserved current: 
\be  \label{GaGE}
 \p_{\mu} \whsF^{\mu\nu }_{a}  = \whmJ_{a}^{\; \nu}   ,\qquad \p_{\nu} \whmJ_{a}^{\; \nu} = 0, 
\ee
where the field strength $\whsF^{\mu\nu }_{a}$ and current $\whmJ_{a}^{\; \nu}$ are defined explicitly in Eqs.(\ref{GGFS}) and (\ref{GC}), respectively. Obviously, when applying the above gauge-type gravitational equation of motion for gravigauge field to the whole energy-momentum tensor given in Eq.(\ref{EMTensor2}) (or Eq.(\ref{EMTensor})), we arrive at zero energy-momentum tensor :
\be
\wh{\cT}_{\mu}^{\; \, \nu} & = &  \bM_{\kappa}^2 \chi_{\mu}^{\; a} (\whmJ_{a}^{\; \, \nu} - \p_{\rho} \whsF^{\rho\nu }_{a} ) =0 ,
\ee
which introduces a stringent constraint that goes beyond the ordinary conservation law of the energy-momentum tensor derived from classical and quantum field theories in the absence of gravitational interactions.

Based on the above observation, we arrive at a profound statement in GQFT that extends beyond Noether's theorem on translational invariance. It states that: a fundamental theory describing the basic constituents of matter and all fundamental interactions among them possesses a zero energy-momentum tensor due to translational invariance within the framework of GQFT. This general theorem on translational invariance may be referred to as the cancellation law of the energy-momentum tensor in GQFT, which fundamentally generalizes the conservation law of the energy-momentum tensor in QFT.

To gain an intuitive understanding of the cancellation law of the energy-momentum tensor, let us analyze the couplings of the gravigauge field and look at the sources of gravitational interactions. It is observed that all fundamental fields with kinetic motion, as well as all gauge fields as vector fields in Minkowski spacetime, experience gravitational interactions through their couplings with the inverse gravigauge field $\chih_{a}^{\; \mu}$. These couplings provide contributions to the energy-momentum tensor that are opposite in sign to those of the gravigauge field. This results in a complete cancellation, leading to a zero energy-momentum tensor for the fundamental theory constructed within the framework of GQFT, which is grounded in the entirety of Minkowski spacetime.

\section{ Equivalence between Zero Energy Momentum Tensor Theorem and General Geometric-type Gravitational Equation}

In the previous section, we demonstrated the equivalence between the gauge-type gravitational equation and the zero energy-momentum tensor. Here, we further explore a potential relationship between the zero energy-momentum tensor and the Einstein equation in GR. To do so, it is useful to verify that the quadratic term of the gravigauge field strength $\sF_{\mu\nu}^{a}$ in the action of the general theory of QED, as presented in Eq. (\ref{actionGQED2}), is equivalent to the Einstein-Hilbert action up to a total derivative. Specifically,
\be \label{GGGR}
& & \frac{1}{4} \chi\, \tchi_{aa'}^{\mu\nu \mu'\nu'} \sF_{\mu\nu}^{a} \sF_{\mu'\nu'}^{a'} \equiv \chi\, R
- 2 \p_{\mu} (\chi \chih^{\mu\rho} \chih_{a}^{\;\sigma} \sF_{\rho\sigma}^{a} ) , 
\ee
with $R$ the Ricci curvature scalar. In general, we have the following identities and relations:
\be \label{GGGI}
 & & R \equiv  \chih_{b}^{\; \mu} \chih_{a}^{\; \nu} R_{\mu\nu}^{ab}  \equiv \chih^{\mu\sigma} \chih^{\nu\rho} R_{\mu\nu\rho\sigma} \equiv \chih^{\mu\sigma} R_{\mu\sigma} , \nn \\
 & & R_{\mu\nu}^{ab} = \p_{\mu}\mOm_{\nu}^{ab} - \p_{\nu}\mOm_{\mu}^{ab} + \mOm_{\mu c}^{a} \mOm_{\nu}^{cb} - \mOm_{\nu c}^{a} \mOm_{\mu}^{cb} , \nn \\
 & & R_{\mu\nu\sigma}^{\;\rho}(x)  = \p_{\mu} \Gamma_{\nu\sigma}^{\rho} - \p_{\nu} \Gamma_{\mu\sigma}^{\rho}  + \Gamma_{\mu\lambda}^{\rho} \Gamma_{\nu\sigma}^{\lambda}  - \Gamma_{\nu\lambda}^{\rho} \Gamma_{\mu\sigma}^{\lambda} ,
\ee
where $R_{\mu\nu}^{ab}$ is defined as the field strength of spin gravigauge field $\mOm_{\mu}^{ab}$, and $R_{\mu\nu\rho\sigma}\equiv \chi_{\rho\lambda}R_{\mu\nu\sigma}^{\;\lambda}$ is Riemann curvature tensor with $R_{\mu\sigma}$ the Ricci curvature tensor in geometry. $\Gamma_{\mu\sigma}^{\rho}(x)$ is the so-called affine connection (or Christoffel symbol) defined by:
\be \label{RMC3}
& & \Gamma_{\mu\sigma}^{\rho}(x)  \equiv  \chih_{a}^{\;\; \rho} \p_{\mu} \chi_{\sigma}^{\;\; a} +  \chih_{a}^{\;\; \rho}   \mOm_{\mu\1 b}^{a} \chi_{\sigma}^{\;\;b} , \nn \\
& & \quad \quad \quad = \frac{1}{2}\chih^{\rho\lambda} (\p_{\mu} \chi_{\lambda\sigma} + \p_{\sigma} \chi_{\lambda\mu} - \p_{\lambda}\chi_{\mu\sigma} ) =\Gamma_{\sigma\mu}^{\rho}.
 \ee

In light of the above relations and identities, the zero energy momentum tensor can be rewritten into the following form: 
\be
\cT_{\mu\nu}  & \equiv & \wh{\cT}_{\mu}^{\; \, \rho} \chi_{\rho\nu} =  \chih [  8\pi G_{\kappa} \mT_{\mu\nu} - (R_{\mu\nu} - \frac{1}{2} \chi_{\mu\nu} R  +  \gamma_W \cG_{\mu\nu} )  ] / 8\pi G_{\kappa}  = 0 ,
\ee
with the definitions:
\be \label{CC}
\mT_{\mu\nu} & = &  \frac{1}{2} ( \chi_{\mu\nu} \chih_{a}^{\; \rho}   - \eta_{\mu}^{\; \rho}  \chi_{\nu a} ) ( \bar{\psi} \gamma^{a} i\cD_{\rho} \psi + H.c.) - \chi_{\mu\nu} m  \bar{\psi}\psi \nn \\
& + & ( \eta_{\mu}^{\; \rho}\eta_{\nu}^{\; \sigma} - \frac{1}{4} \chi_{\mu\nu}  \chih^{\rho\sigma} )\chih^{\rho'\sigma'}  ( F_{\rho\rho'}F_{\sigma\sigma'} + \cF_{\rho\rho'}^{ab} \cF_{\sigma\sigma' ab} )  , \nn \\
\cG_{\mu\nu} & = &  \frac{1}{2} \chih\, \chi_{\mu}^{\;\, a}\cD_{\rho}(\widehat{\cG}_{a}^{\rho\sigma})\chi_{\sigma\nu} +  \frac{1}{2}  (\eta_{\mu}^{\; \rho}\chi_{\lambda\nu } -  \frac{1}{4} \chi_{\mu\nu}\eta_{\lambda}^{\; \rho} ) \chih \cG_{\rho\sigma}^{a} \widehat{\cG}^{\lambda\sigma}_{a}  . 
\ee

Clearly, the zero energy-momentum tensor in GQFT leads to the general geometric-type gravitational equation:
 \be \label{GGE}
 R_{\mu\nu} - \frac{1}{2} \chi_{\mu\nu} R  +  \gamma_W \cG_{\mu\nu} = 8\pi G_{\kappa} \mT_{\mu\nu} ,
 \ee
which provides a comprehensive description of gravidynamics within the framework of GQFT. Unlike in GR,  the spin gauge invariant tensors $\mT_{\mu\nu}$ and $\cG_{\mu\nu}$ are generally not symmetric in GQFT. Specifically, $\mT_{\mu\nu} \neq \mT_{\nu\mu}$ and $\cG_{\mu\nu} \neq \cG_{\nu\mu}$. This allows the above general geometric-type gravitational equation to be decomposed into symmetric and antisymmetric components:
 \be \label{GGE2}
 R_{\mu\nu} -  \frac{1}{2}\chi_{\mu\nu} R  + \gamma_W \cG_{(\mu\nu)} & = & 8\pi G_{\kappa} \mT_{(\mu\nu)} , \nn \\
\gamma_W \cG_{[\mu\nu]} & = & 8\pi G_{\kappa} \mT_{[\mu\nu]} , 
\ee
with the symmetric tensors $\mT_{(\mu\nu)}$ and $\cG_{(\mu\nu)}$, and antisymmetric tensors $\mT_{[\mu\nu]}$ and $\cG_{[\mu\nu]}$ defined as follows:
\be
 \mT_{\mu\nu}  & \equiv & \mT_{(\mu\nu)} + \mT_{[\mu\nu]}, \quad \cG_{\mu\nu}   \equiv \cG_{(\mu\nu)} + \cG_{[\mu\nu]}, \nn \\
\mT_{(\mu\nu)}  & \equiv &  \frac{1}{2} (\mT_{\mu\nu} + \mT_{\nu\mu} ), \;\; \mT_{[\mu\nu]}  \equiv  \frac{1}{2} (\mT_{\mu\nu} - \mT_{\nu\mu} ), \nn \\
\cG_{(\mu\nu)}  & \equiv &  \frac{1}{2} (\cG_{\mu\nu} + \cG_{\nu\mu} ), \;\; \; \cG_{[\mu\nu]}  \equiv  \frac{1}{2} (\cG_{\mu\nu} - \cG_{\nu\mu} ).
\ee 

The symmetric equation leads to a generalized Einstein equation, while the antisymmetric equation provides an additional constraint. This additional equation arises due to the presence of the spin-related gravigauge field as the fundamental gravitational field and the spin gauge field, both of which originate from the spin gauge symmetry SP(1,3).

\section{Flowing and Entire unitary gauges and the basic symmetry in GQFT }

The gauge-type gravitational equation in Eq. (\ref{GaGE}) demonstrates that the gravitational interaction, characterized by the gravigauge field in GQFT, emerges as a gauge-type theory within the framework of QFT. In this framework, the coordinate spacetime remains a flat Minkowski spacetime with global Poincar\'e group symmetry. Additionally, it is noteworthy that although GQFT possesses the spin gauge symmetry SP(1,3), its spin gauge field $\cA_{\mu}^{ab}$ allows to have a mass term $M_{\cA}$, as explicitly shown in the action Eq. (\ref{actionGQED2}) and in the equation of motion for the massive spin gauge field in Eq. (\ref{EoMSG}). This is possible because the spin gravigauge field $\mOm_{\mu}^{ab}$, characterized by the dual gravigauge fields $\chi_{\mu}^{\; a}$ and $\chih_{a}^{\; \mu}$ as shown in Eq. (\ref{SGGF}), and the spin gauge field $g_s\cA_{\mu}^{ab}$ share the same transformation properties under the spin gauge symmetry SP(1,3). The spin gauge invariance for the massive spin gauge field can also be verified from the relation given in Eq. (\ref{MSG}), owing to the explicit gauge-covariant field strength $\cG_{\mu\nu}^{a}$ of the gravigauge field $\chi_{\mu}^{\; a}$. This feature distinguishes intrinsic spin gauge symmetry in GQFT from ordinary internal gauge symmetries, where a massive gauge field usually breaks the corresponding gauge symmetry. Instead, in GQFT, this property for the spin gauge symmetry is attributed to the associated gravigauge field, which behaves as a Goldstone boson, as indicated in Eq. (\ref{MSG}).

In general, any internal gauge symmetry introduces redundant degrees of freedom, which must be eliminated by imposing an appropriate gauge-fixing condition. For the spin gauge symmetry SP(1,3) of the spinor field, the situation becomes particularly unique. It not only necessitates the introduction of a corresponding spin gauge field $\cA_{\mu}^{ab}$ but also unavoidably requires an invertible bi-covariant vector field $\chi_{\mu}^{\;\; a}$, along with its dual field $\chih_{a}^{\; \mu}$, referred to as the gravigauge field. This gravigauge field contains sixteen degrees of freedom, which includes six additional degrees of freedom compared to the composite symmetric gravimetric field $\chi_{\mu\nu}= \chi_{\mu}^{\; a} \chi_{\mu}^{\; b} \eta_{ab}$ in GR. These extra degrees of freedom reflect the equivalence classes of the spin gauge symmetry SP(1,3), allowing us to eliminate the redundant degrees of freedom by applying a gauge prescription to the spin gauge symmetry.

A simple gauge prescription can be realized by performing a special spin gauge transformation, $S(\tilde{\Lambda})$, in the spinor representation, where the $\gamma$-matrix and the gravigauge field transform as follows:
\be
S^{-1}(\tilde{\Lambda}) \gamma^a S(\tilde{\Lambda}) = \tilde{\Lambda}^{a}_{\;\, b}(x) \gamma^b, \quad 
\chi_{\mu a }(x) \to \chi_{\mu a}(x) \tilde{\Lambda}^{a}_{\; \, b}(x)  =  \tchi_{\mu b}(x) ,
\ee
which allows the gravigauge field, behaving as a Goldstone boson, to be transmuted into a symmetric form,
\be
& &  \tchi_{\mu a}(x)  =  \tchi_{a\mu}(x) . 
\ee 
Such a spin gauge transformation provides a natural gauge-fixing prescription for the spin gauge symmetry 
SP(1,3), ensuring that the symmetric gravigauge field possesses the same number of degrees of freedom as the symmetric gravimetric field $\chi_{\mu\nu}(x)$. Consequently, the total independent degrees of freedom in the theory remain unchanged, as the extra degrees of freedom in the gravigauge field are transmuted into the spin gauge field $\cA_{\mu}^{ab}(x)$, which becomes a massive gauge field.

Such a gauge prescription is referred to as the flowing unitary gauge, as it is valid only within a local coordinate system that varies from point to point in Minkowski spacetime. This arises from the fact that the action presented in Eq. (\ref{actionGQED2}) can be shown to possess a hidden general linear group symmetry GL(1,3,R) in coordinate spacetime. This observation can be fully understood from the action presented in Eq. (\ref{actionGQED}), where the action is constructed within the framework of gravigauge spacetime. This construction ensures that the action is independent of the choice of coordinate systems and naturally generates the general linear group symmetry GL(1,3,R) as a hidden local symmetry. Consequently, the action in Eq. (\ref{actionGQED}) actually exhibits a maximal joint symmetry:
\be \label{JGLS}
G_S = \mbox{GL(}1,3,\mbox{R)} \Join SP(1, 3),
\ee
which indicates that the global Poincar\'e group symmetry PO(1,3), which underlies GQFT, is automatically extended to a local group symmetry GL(1,3,R) when the gauge invariance principle is applied to construct an action in gravigauge spacetime. It is well known that the group symmetry GL(1,3,R) forms the foundation of Einstein's GR and governs gravitational interactions in curved Riemannian spacetime. Consequently, it is manifest that GQFT, developed based on the intrinsic spin gauge symmetry within the framework of QFT, naturally leads to the general geometric-type gravitational equation presented in Eq. (\ref{GGE}). This equation adheres to the principle of general coordinate covariance under the transformation of the general linear group symmetry

It is evident that when performing a local transformation of the group symmetry GL(1,3,R), the symmetric gravigauge field resulting from the flowing unitary gauge no longer remains symmetric. To preserve the symmetric gravigauge field, $\tchi_{\mu a}(x)  =  \tchi_{a\mu}(x)$, in the flowing unitary gauge, it is necessary to first apply a gauge fixing for the general linear group symmetry GL(1,3,R). A gauge prescription can be implemented by requiring the spin gauge covariant derivative of the gravigauge field to vanish,
\be
\cD^{\mu} \chi_{\mu}^{\; a} \equiv \chih^{\mu\nu}(\p_{\nu} \chi_{\mu}^{\; a} + \cA_{\nu b}^{a} \chi_{\mu}^{\; b} )  =0 ,
\ee 
which fixes the general linear group symmetry GL(1,3,R) while simultaneously allowing a special spin gauge transformation to preserve the symmetric gravigauge field. Within such a gauge prescription, the flowing unitary gauge can be extended to an entire unitary gauge.
 
In the entire unitary gauge, the action possesses a global associated symmetry:
\be \label{GAS}
G_S = P^{1,3} \ltimes \mbox{SO}(1,3) \adjoin \mbox{SP}(1, 3)\equiv PO(1,3) \adjoin SP(1, 3),
\ee
which becomes the fundamental symmetry after applying the appropriate gauge-fixing conditions to eliminate all redundant degrees of freedom.

\section{Distinct Foundations of GQFT and Poincar\'e gauge theory  }

It is demonstrated that the global Poincar\'e symmetry serves as a foundational symmetry not only in QFT but also in GQFT within the entire unitary gauge. This naturally motivates a generalization of the global Poincar\'e symmetry to a local symmetry, a concept formally realized in the Poincar\'e gauge theory (PGT) of gravity. PGT was proposed to encompass Einstein’s general relativity, the Einstein-Cartan theory, and teleparallel gravity. However, while both GQFT and PGT originate from the gauge invariance principle, their conceptual foundations differ fundamentally.

The essential notion of GQFT lies in reconciling the foundational frameworks of GR and QFT through a guiding fundamental principle: the laws of nature are dictated by the intrinsic properties of elementary particles as quantum fields. This necessitates a clear distinction between intrinsic symmetries, determined by the quantum numbers of elementary particles, and external symmetries, characterized through their motion within the coordinate-dependent Minkowski spacetime. This critical distinction leads to the emergence of two independent symmetries in GQFT: the global Poincaré symmetry PO(1,3) in Minkowski spacetime and the intrinsic spin gauge symmetry SP(1,3) in the Hilbert space of Dirac spinor fields. These symmetries combine to form a unified joint symmetry, denoted PO(1,3)$\Join$SP(1,3). This crucial distinction leads the global Poincar\'e symmetry PO(1,3) in Minkowski spacetime of coordinates and the intrinsic spin gauge symmetry SP(1,3) in Hilbert space of Dirac spinor field to act as a joint symmetry PO(1,3)$\Join$SP(1,3) in GQFT. To preserve this joint symmetry, an invertible bi-covariant vector field, denoted $\chih_{a}^{\; \mu}(x)$, must be introduced alongside the spin gauge field $\cA_{\mu}^{ab}$. 
Consequently, the spin-related bi-covariant vector field $\chi_{\mu}^{\; \, a}(x)$ as a dual field of $\chih_{a}^{\; \mu}(x)$, termed the gravigauge field, emerges as the fundamental gravitational field. Crucially, as shown in Eq. (\ref{GT}), both the gravigauge field $\chi_{\mu}^{\; a}(x)$ and its dual field $\chih_{a}^{\; \mu}(x)$ must transform homogeneously as bi-covariant vector fields under both spin gauge symmetry SP(1,3) and global Lorentz symmetry SO(1,3) to maintain the joint symmetry.

As for PGT, it was developed in analogy to GR, which was established by generalizing special relativity (SR). PGT applies the gauge principle to the external Poincaré group symmetry PO(1,3)\cite{GGT2}, which is the fundamental symmetry group governing motions in Minkowski spacetime within SR. The gauging of the Lorentz subgroup SO(1,3) was first explored in early studies\cite{GGT1,GGT3} by extending the framework of conventional internal gauge theories\cite{GT1,GT2}. PGT introduces two gauge potentials: the tetrad (or vierbein) field $e_{\mu}^{\; m}$ and the Lorentz connection $\Gamma_{\mu}^{mn} = - \Gamma_{\mu}^{nm}$, which respectively facilitate the gauge symmetries of the translation subgroup $P^{1,3}$ and the Lorentz subgroup SO(1,3).

In general, PGT extends GR by yielding two field equations of gravity: one for the tetrad field and another for the torsion field. The general structure of PGT has been extensively analyzed in references\cite{PGT1,PGT2}. Notably, the Einstein-Cartan theory\cite{EC} is characterized by an action that depends solely on the curvature scalar of Riemann–Cartan spacetime. On the other hand, the translation gauge theory of gravity, or teleparallel gravity, with zero curvature can also be made compatible with GR. Both cases are regarded as degenerate forms of PGT. It has been shown in ref.\cite{Tele1} that there exists a unique, consistent choice of the teleparallel action that is equivalent to GR. However, this equivalence no longer holds when the theory is coupled to a Dirac field, as demonstrated in refs.\cite{Tele2,Tele3}.

To make an explicit analysis, let us begin with the following covariant derivative:
\be
& & D_{\mu} \equiv \p_{\mu} + \varGamma_{\mu},  \nn \\
& & \varGamma_{\mu} \equiv \varGamma_{\mu}^{M} \cP_M = e_{\mu}^{\; m} P_m + \Gamma_{\mu}^{m n} J_{m n} ,
\ee
with indices $m, n =0,1,2,3$ denoting Lorentz vectors, raised and lowered by the constant metric tensors $\eta^{m n}$($\eta_{mn}$) =$\diag.$(1,-1,-1,-1). Here, $\varGamma_{\mu}^{M} \equiv ( e_{\mu}^{\; m}, \Gamma_{\mu}^{m n} )$ and $\cP_M\equiv ( P_m, J_{m n})$ represent the gauge potentials and group generators of external Poincar\'e gauge symmetry PO(1,3), respectively. 
The group generators $\cP_M$ satisfy the following Lie algebra of PO(1,3):
\be
& &[J_{mn}, J_{m'n'} ] =  J_{mn'} \eta_{nm'} -  J_{nn'} \eta_{mm'}  + J_{nm'}\eta_{mn'} - J_{mm'}\eta_{nn'}, \nn \\
& & [J_{mn}, P_{l} ] =  P_{m} \eta_{nl}  - P_{n}\eta_{ml}, \quad [P_{m}, P_{n}] = 0 .
\ee
Under the gauge transformations of PO(1,3) group symmetry, the gauge fields transform as follows:
\be \label{GTPO}
& & \delta \varGamma_{\mu} = D_{\mu} \varrho = \p_{\mu} \varrho + [\varGamma, \varrho] , \nn \\
& & \delta e_{\mu}^{\; m} = \varrho^{m}_{\; n} e_{\mu}^{\; n} +  \p_{\mu}  \varrho^{m} + \Gamma_{\mu n}^{m} \varrho^{n} , \nn \\
& & \delta \Gamma_{\mu}^{m n} = \p_{\mu}\varrho^{m n} + \Gamma_{\mu l}^{m} \varrho^{ln} - \Gamma_{\mu l}^{n} \varrho^{lm} ,
\ee
with $\varrho = \varrho^{m} P_{m} + \varrho^{mn} J_{mn}$ an infinitesimal gauge parameter.

By considering the following quadratic form:
\be
& & \langle J_{mn}\; J_{m'n'} \rangle = \frac{1}{2} (\eta_{mm'}\eta_{nn'} -  \eta_{nm'}\eta_{mn'} ), \nn \\
& & \langle P_{m}\; P_n \rangle = \lambda_{\kappa} \eta_{mn} ,\quad \langle J_{mn}\; P_{l} \rangle = 0, 
\ee
a general structure for the PGT Lagrangian can be constructed. This approach is analogous to the construction of Lagrangians in conventional gauge theories based on internal symmetries, where the Lagrangian is typically built from a quadratic form of the gauge field strength. Specifically, the PGT Lagrangian can be formulated in terms of a quadratic form of the gauge field strength as follows:
\be
\cL_{PGT} = - g^{\mu\mu'}g^{\nu\nu'} \frac{1}{4g_{\kappa}^2} ( \cR_{\mu\nu}^{mn}\cR_{\mu'\nu' mn} + \lambda_{\kappa} \cT_{\mu\nu}^{m} \cT_{\mu'\nu' m}) ,
\ee 
with the field strengths defined by,
\be
& & \cR_{\mu\nu}^{mn} = \p_{\mu}\Gamma_{\nu}^{mn} -  \p_{\nu}\Gamma_{\mu}^{mn} + \Gamma_{\mu m'}^{m} \Gamma_{\nu}^{m' n} - \Gamma_{\nu m'}^{m} \Gamma_{\mu}^{m' n}, \nn \\
& & \cT_{\mu\nu}^{m} = \p_{\mu}e_{\nu}^{m} -  \p_{\nu}e_{\mu}^{m} + \Gamma_{\mu n}^{m}e_{\nu}^{n} - \Gamma_{\nu n}^{m} e_{\mu}^{n} .
\ee

In the above Lagrangian, a contravariant tensor $g^{\mu\nu}$ is adopted to obey the principle of general coordinate invariance under transformations of the general linear group symmetry GL(1,3,R), which serves as the foundation of GR. This implies that the tensor field $g^{\mu\nu}$ in PGT is introduced as an independent and invertible tensor by principle. Consequently, its dual covariant metric tensor $g_{\mu\nu}$ also becomes independent. In PGT, the vierbein $e_{\mu}^{\; m}$, which acts as the gauge field for the translation subgroup of the Poincar\'e group PO(1,3), does not transform homogeneously under the gauge transformation of PO(1,3) as given in Eq. (\ref{GTPO}). As a result, it is not possible to directly construct a gauge-invariant metric tensor using a bilinear form of the vierbein, namely, 
\be
g_{\mu\nu} \neq e_{\mu}^{\; m}e_{\nu}^{\; n}\eta_{m n} ,
\ee 
which suggests that Poincar\'e gauge theory is not straightforwardly related to GR, which is characterized by the metric tensor $g_{\mu\nu}$. This contrasts with the conventional approach, where the metric tensor is typically expressed as a bilinear form of the vierbein, $g_{\mu\nu} = e_{\mu}^{\; m}e_{\nu}^{\; n}\eta_{m n}$. 

Regarding the coupling of the vierbein $e_{\mu}^{\; m}$ to the Dirac spinor field, when considering the Lorentz symmetry group SO(1,3) that is isomorphic to the spin symmetry group SP(1,3), the coupling takes the following general form:
\be
\cL ( \psi, e_{\mu}^{\; m}, g_{\mu\nu} ) = g^{\mu\nu} \bar{\psi}\gamma_{m} e_{\mu}^{\; m} iD_{\nu} \psi ,
\ee
which is not well-defined due to the same reason mentioned above, the vierbein $e_{\mu}^{\; m}$, as a gauge field in PGT, does not transform homogeneously. Consequently, the Lagrangian is not gauge-invariant under transformations of the Poincar\'e gauge symmetry.

It becomes evident that PGT does not appear to be a well-formulated gauge theory of gravity, despite being motivated by the gauging of the external symmetry of motions in Minkowski spacetime within the framework of SR.

Unlike PGT, where the vierbein $e_{\mu}^{\; m}$, as a gauge field of the translation subgroup of the external Poincar\'e gauge symmetry group, is unsuitable for defining the metric tensor field, in GQFT, a spin-related invertible gravigauge field $\chi_{\mu}^{\; a}$ emerges as a bi-covariant vector field associated with the spin gauge symmetry SP(1,3) of the Dirac spinor field. This gravigauge field behaves as a Goldstone boson. Its field strength $\sF_{cd}^{a}$ (defined in Eq. (\ref{NCR})) is governed by the gravitization equation (given in Eq. (\ref{GE})), which is derived from a constraint equation as $\sF_{cd}^{a}$ appears to be an auxiliary field in gravigauge spacetime. This leads to the observation that the gravitational effect in gravigauge spacetime arises from the non-commutative nature of the gravigauge derivative operator and is governed by the collective dynamics of all gauge fields through the gravitization equation. 

Consequently, we would like to emphasize that, within the framework of GQFT, gravidynamics is described by the gauge-type gravitational equation derived from the equation of motion for the gravigauge field, as shown in Eq. (\ref{GaGE}). Notably, the mass-like term of the spin gauge field, represented by the relation in Eq. (\ref{MSG}), is essential for ensuring the consistency of the theory. This is particularly relevant when considering an appropriate gauge prescription for the flowing unitary gauge of spin gauge symmetry, as discussed in the previous section. Furthermore, this term becomes indispensable when the gravigauge field serves as the fundamental gravitational field coupled to the Dirac spinor field, as evidenced from the geometric-type gravitational equation in Eq.(\ref{GGE2}). Through the gauge-geometry correspondence demonstrated in Eq. (\ref{GGGR}), this establishes an equivalence between the action of gravigauge field and the Einstein-Hilbert action of GR.

\section{GQFT in three-dimensional spacetime and its connection to Witten's perspective on three-dimensional gravity}

As GQFT is founded on the fundamental principle that the laws of nature are governed by the intrinsic properties of quantum fields as the basic constituents of matter, it has been demonstrated in previous references\cite{FHUFT1,FHUFT2,FHUFT} that spacetime dimensions are determined by the independent degrees of freedom of the qubit-spinor field guided by the principle of maximal coherence motion. Furthermore, fundamental interactions are governed by intrinsic symmetries based on the gauge invariance principle. For our current purpose, we will focus on a qubit-spinor field with two real components, which has been shown to exhibit maximal motion in three dimensional spacetime. 

To be explicit, let us start with the action of freely moving two-component qubit-spinor field,
\be \label{3Daction1}
 \cS_{3D} & = & \int d^3x \, \{ \frac{1}{2} \bar{\psi}(x) \delta_a^{\;\mu} \gamma^a i\partial_{\mu} \psi(x)  -  \frac{1}{2} m_3 \bar{\psi}(x) \tga  \psi(x) \} ,
 \ee
with the $\gamma$-matrices $\gamma^a$ (a=0,1,2) and qubit-spinor field $\psi(x)$ defined as follows:
\be \label{UQSF}
& & \psi(x) \equiv \binom{\psi_{0}(x)}{\psi_{1}(x)} , \quad  \bar{\psi}(x)  = \psi^{\dagger}(x) \gamma^0,   \nn \\
&&\gamma^0 = \sigma_2 , \quad \gamma^1  = i \sigma_1 , \quad   \gamma^2 = - i \sigma_3 , \quad \tga = \sigma_0, 
\ee
where the components $\psi_0(x)$ and $\psi_1(x) $ are two real functions, $m_3$ denotes the mass of qubit-spinor field in three-dimensional spacetime, and all $\gamma$-matrices are imaginary ones. It is easy to check that the above action has an associated symmetry, 
\be
G_S = P^{1,2}\ltimes SO(1,2) \adjoin SP(1,2) \equiv PO(1,2) \adjoin SP(1,2), 
\ee
with PO(1,2) representing Poincar\'e group symmetry (inhomogeneous Lorentz symmetry) in three dimensional Minkowski spacetime, and SP(1,2) denoting spin symmetry acting on the qubit-spinor field. 

To exhibit a maximal symmetry, it is useful to express the qubit-spinor field in a self-conjugated chiral representation as follows:
\be
\psi_{-} = \binom{0}{\psi(x)} \quad \mbox{or} \quad  \psi_{+} = \binom{\psi(x)}{0},
\ee  
which have the properties:
\be \label{CM}
& & \gamma_5 \psi_{-} = - \psi_{-}, \quad  \gamma_5 \psi_{+} = - \psi_{+}, \nn \\
& & \psi_{+} \equiv  \psi_{-}^{c_d} = \cC_d^{-1} \psi_{-} \cC_d = C_d \psi_{-}, \nn \\
& & \gamma_5 = \sigma_3\otimes \sigma_0, \quad C_d = \sigma_1 \otimes \sigma_0, 
\ee
where the self-conjugated chiral qubit-spinor fields $\psi_{-}$ and $\psi_{+}$ possess negative and positive chirality, respectively, and they are related via the chiral duality operation $\cC_d$. 

The physical laws are independent of the specific choice between the two chirality representations, $\psi_{-}$ and $\psi_{+}$, as these representations should be physically equivalent. This leads to the principle of chirality independence, which requires that the resulting theory must remain invariant under the chiral duality operation $\cC_d$.

By applying for the equivalence of the chirality representations $\psi_{-}$ and $\psi_{+}$, the above action can be reformulated into the following form, which is invariant under chiral duality:
\be  \label{3Daction2}
\cS_{3D} & = &\int d^3x  \frac{1}{4} \{ \bar{\psi}_{-} \gamma^{a} \gamma_{-} \delta_{a}^{\; \mu} i\p_{\mu} \psi_{-}  
- m_3 \bar{\psi}_{-} \gamma^{3}\gamma_{-} \psi_{-}  \nn \\
& + &  \bar{\psi}_{+} \gamma^{a} \gamma_{+} \delta_{a}^{\; \mu} i\p_{\mu} \psi_{+} + m_3 \bar{\psi}_{+} \gamma^{3} \gamma_{+} \psi_{+} \} ,
\ee
with the $\gamma$-matrices $\gamma^{a}$ (a=0,1,2,3) defined explicitly as follows: 
\be \label{GM2}
& &  \gamma^0= \sigma_1 \otimes \sigma_2,  \nn \\
& & \gamma^1 = i \sigma_1\otimes \sigma_1, \nn \\
& & \gamma^2 = -i\sigma_1\otimes \sigma_3, \nn \\
& & \gamma^3 = - i\sigma_2\otimes \sigma_0,  \nn \\
& & \gamma_{\mp} = \frac{1}{2} ( 1 \mp \gamma_5 ) .
\ee

The chiral duality, characterized by $C_d^2 = 1$, exhibits a $Z_2$ discrete symmetry in the above action. Additionally, it can be demonstrated that this action possesses an enlarged associated symmetry,
\be
G_S = P^{1,2}\ltimes SO(1,2) \adjoin SP(1,2) \rtimes W^{1,2} \times Z_2 \equiv PO(1,2) \adjoin WS(1,2) \times Z_2,
\ee
where WS(1,2)=SP(1,2) $\rtimes$ W$^{1,2}$ represents an inhomogeneous spin symmetry with W$^{1,2}$ denoting a chirality boost-spin symmetry. The group generators $\Sigma_{A-} \equiv (\Sigma_{ab}, \Sigma_{a -} )$ of WS(1,2) are explicitly given by,
\be
\Sigma_{ab} = \frac{i}{4} [ \gamma_a , \gamma_b ], \quad \Sigma_{a - } = \gamma_a \gamma_{-}, 
\ee
which satisfies the following Lie algebra relations:
\be
& & [\Sigma_{ab}, \Sigma_{cd} ] = i ( \Sigma_{ad} \eta_{bc} - \Sigma_{bd} \eta_{ac} +  \Sigma_{bc} \eta_{ad} - \Sigma_{ac} \eta_{bd} ) , \nn \\
& & [\Sigma_{ab}, \Sigma_{c -} ] = i (  \Sigma_{a -} \eta_{bc} - \Sigma_{b -} \eta_{ac} ), \quad  [\Sigma_{a -}, \Sigma_{b -} ] = 0 .
\ee

It is important to note that, unlike the inhomogeneous Lorentz symmetry (Poincar\'e symmetry), the translation-like generators of subgroup W$^{1,2}$ are nilpotent, i.e., $\Sigma_{a -}^2 = 0$. The chiral duality implies that the group generators for the positive chirality representation $\psi_{+}$ can be derived though the chiral duality operation as follows:
\be
\Sigma_{A+}  = \cC_{d}^{-1} \Sigma_{A-}\cC_{d} = (\Sigma_{ab}, \Sigma_{a +} ).
\ee

Let us now consider the interactions of the qubit-spinor field by applying the gauge invariance principle. Specifically, the intrinsic inhomogeneous spin symmetry WS(1,2) should be gauged as a local symmetry. This necessitates the introduction of a gauge field $\bscA_{\mu}^{\mp}$, associated with dual gravigauge field $\chih_{a}^{\; \mu}$, to ensure the spin gauge symmetry WS(1,2). Concretely, the ordinary derivative operator must be replaced by a covariant derivative operator, as follows:
\be
& & \delta_{a}^{\; \mu} i\p_{\mu} \to \chih_{a}^{\; \mu} i\cD_{\mu}^{\mp} , \quad i\cD_{\mu}^{\mp}  = i\p_{\mu} + \mu_s \bscA_{\mu}^{\mp} \equiv i\cD_{\mu} + \cW_{\mu}^{\mp} ,  \nn \\
& & \bscA_{\mu}^{\mp} \equiv \cA_{\mu}^{A} \Sigma_{A \mp} \equiv \cA_{\mu} + \cW_{\mu}^{\mp} \equiv \cA_{\mu}^{ab}\frac{1}{2}\Sigma_{ab} +  \cW_{\mu}^{\; a} \Sigma_{a\mp} ,
\ee  
where a chirality boost-spin gauge field $\cW_{\mu}^{a}$ is introduced alongside the spin gauge field $\cA_{\mu}^{ab}$, and $\mu_s$ represents a gauge coupling constant. Both fields and the coupling constant have a dimension of $(mass)^{1/2}$. Their corresponding field strengths are expressed as follows:
\be
& & \bscF_{\mu\nu}^{\mp} = \p_{\mu} \bscA_{\nu}^{\mp} - \p_{\nu} \bscA_{\mu}^{\mp} + i \mu_s [\bscA_{\mu}^{\mp} , \bscA_{\nu}^{\mp} ] \equiv  \cF_{\mu\nu} + \cF_{\mu\nu}^{\mp} , \nn \\
& & \cF_{\mu\nu}  \equiv \cF_{\mu\nu}^{ab} \frac{1}{2} \Sigma_{ab} , \quad \cF_{\mu\nu}^{\mp} \equiv  \cF_{\mu\nu}^{ a} \Sigma_{a \mp} , \nn \\
& & \cF_{\mu\nu}^{ab} =  \p_{\mu} \cA_{\nu}^{ab} - \p_{\nu} \cA_{\mu}^{ab} +   \mu_s (\cA_{\mu c}^{a} \cA_{\nu}^{cb} - \cA_{\nu c}^{a} \cA_{\mu}^{cb} ) , \nn \\
& & \cF_{\mu\nu}^{a} =  \p_{\mu} \cW_{\nu}^{\; a} - \p_{\nu} \cW_{\mu}^{\; a} +   \mu_s (\cA_{\mu b}^{a} \cW_{\nu}^{\; b} - \cA_{\nu b}^{a} \cW_{\mu}^{\; b} ) .
\ee
It can be demonstrated that under the transformations of the inhomogeneous spin gauge symmetry WS(1,2), the gauge fields exhibit the following transformation properties:
\be \label{GTPO}
& & \delta \bscA_{\mu}^{\mp} = D_{\mu} \varpi^{\mp} = \p_{\mu} \varpi^{\mp} + [\bscA_{\mu}^{\mp}, \varpi^{\mp}] , \nn \\
& & \delta \cA_{\mu}^{a b} = \p_{\mu}\varpi^{a b} + \cA_{\mu c}^{a} \varpi^{ c b} - \cA_{\mu c}^{b} \varpi^{c b} , \nn \\
& & \delta \cW_{\mu}^{\; a} = \varpi^{a}_{\; b} \cW_{\mu}^{\; b} +  \p_{\mu}  \varpi^{a} + \cA_{\mu b}^{a} \varpi^{b} , \nn \\
& & \varpi^{\mp} =  \varpi^{a b} \Sigma_{ab} /2 + \varpi^{a} \Sigma_{a \mp}, 
\ee
with $ \varpi^{\mp} $ being an infinitesimal gauge parameter.

The corresponding definitions of the covariant derivative operator and the field strength in the spin-related gravigauge spacetime are provided by the following relations:
\be
\hcD_{c}^{\mp} \equiv \chih_{c}^{\; \mu} \cD_{\mu}^{\mp}, \quad \bshcF_{cd}^{\mp} \equiv \chih_{c}^{\; \mu}\chih_{c}^{\; \nu} \bscF_{\mu\nu}^{\mp}.
\ee
Similar to the construction of the general theory of QED in Eq.(\ref{actionGQED2}), the chiral duality invariant action in Eq.(\ref{3Daction2}) can be extended to an inhomogeneous spin gauge invariant action as follows:
\be  \label{3Daction3}
\cS_{3D}  & = & \int [d \zeta] \cL_{3D} \nn \\
& = &\int d^3 \zeta\, \{ \frac{1}{4} [ \bar{\psi}_{-} \gamma^{c} \gamma_{-}  i\hcD_{c}^{-} \psi_{-}  - m_3 \bar{\psi}_{-} \gamma^{3}\gamma_{-} \psi_{-}  \nn \\
& + &  \bar{\psi}_{+} \gamma^{c} \gamma_{+} i\hcD_{c}^{+} \psi_{+} + m_3 \bar{\psi}_{+} \gamma^{3} \gamma_{+} \psi_{+} ] ,\nn \\
& - & \frac{1}{4} \eta^{c c'}\eta^{d d'} \Tr (\bshcF_{cd}^{-}\bshcF_{c'd'}^{-} + \bshcF_{cd}^{+}\bshcF_{c'd'}^{+} ) \nn \\
& + &  \frac{1}{4} \bM_{\kappa} \etat^{cdc'd'}_{aa'} \hsF_{cd}^{a}\hsF_{c'd'}^{a'}  +  M_{\cA}^2 \eta^{cd}  \Tr \bshcA_{c}\bshcA_{d} \nn \\
& + &  \frac{1}{8} \eta^{c c'}\eta^{d d'} \Tr \bshcF_{cd}^{(w-)} \bshcF_{c'd'}^{(w+)}  - \frac{1}{4} m_w^2 \eta^{cd}  \Tr \bshcW_{c}^{(-)}  \bshcW_{d}^{(+)} \} ,
\ee
where we have adopted the following definitions:
\be
& & \bshcA_{c} \equiv \hcA_{c} -  \hmOm_{c}/\mu_s = (\hcA_{c}^{ab} -  \hmOm_{c}^{ab}/\mu_s)\Sigma_{ab}/2, \nn \\
& & \bshcW_{c}^{(\mp)}  \equiv (\hcW_{c}^{\; a} - \hcD_{c}\bs{\zeta}^{a} )  \Sigma_{\mp a} , \; \hcD_{c}\bs{\zeta}^{a} = \eth_{c} \bs{\zeta}^{a}  + \hcA_{c b}^{a}\bs{\zeta}^{b}, \nn \\
& & \bshcF_{cd}^{(w\mp)}  \equiv \hcD_{c} \bshcW_{d}^{(\mp)} - \hcD_{d} \bshcW_{c}^{(\mp)} \equiv (\hcF_{cd}^{a} - \hcF_{cd}^{ab} \bs{\zeta}_{b} ) \Sigma_{\mp a} ,
\ee
and introduced a vector field $\bs{\zeta}^{a}(x)$, valued in three dimensional gravigauge spacetime, to define the chirality boost-spin gauge invariant field $\bshcW_{c}^{(\mp)}$. The vector field $\bs{\zeta}^{a}(x)$ transforms as follows:
\be
\bs{\zeta}^{a}(x) \to \bs{\zeta}^{' a}(x) = \bs{\zeta}^{a}(x) + \varpi^{a}(x) .
\ee
So both $\bshcA_{c}$ and $\bshcW_{c}^{(\mp)}$ become spin gauge covariant fields. 

The translational property of the vector field $\bs{\zeta}^{a}(x)$ enables us to select a gauge transformation of the chirality boost-spin gauge symmetry W$^{1,3}$ to impose an appropriate gauge fixing condition with $\bs{\zeta}^{' a}(x)=0$. This condition is referred to as a unitary chirality boost gauge basis. In this unitary gauge basis, we arrive at the following simplified expressions:
\be \label{CBSGB}
\bs{\zeta}^{a}(x) \to 0, \quad \bshcW_{c}^{(\mp)} \to \hcW_{c}^{\; a} \Sigma_{\mp a}, \;\; \bshcF_{cd}^{(w\mp)} \to \hcF_{cd}^{a}\Sigma_{\mp a} .
\ee

It is easy to check that the trace of group generators has the following quadratic forms:
\be \label{QF}
& & \Tr \Sigma_{A\mp}\Sigma_{C\mp}  = \Tr \Sigma_{ab}\Sigma_{cd} = \eta_{ac}\eta_{bd} -  \eta_{bc}\eta_{ad} , \nn \\
& & \Tr \Sigma_{ab}\Sigma_{c -} = 0, \quad \Tr \Sigma_{a -}\Sigma_{b -} = 0, 
\ee
and the coupling of the chirality boost-spin gauge field to the qubit-spinor field becomes vanishing due to the chirality structure, i.e.:
\be
\bar{\psi}_{\mp} \gamma^{c} \gamma_{\mp}  \hcW_{c}^{\; a}\Sigma_{a\mp} \psi_{\mp} = 0 .
\ee

By projecting the above action, formulated in the gravigauge spacetime, into the framework of GQFT and adopting the chirality boost-spin gauge basis, we arrive at the following action:
\be \label{3Daction4}
\cS_{3D}  & = & \int [d x]\chi \cL_{3D} \nn \\
& = &\int d^3x \, \chi \{ \frac{1}{4} [ \bar{\psi}_{-} \gamma^{a} \gamma_{-} \chih_{a}^{\; \mu}  i\cD_{\mu} \psi_{-}  - m_3 \bar{\psi}_{-} \gamma^{3}\gamma_{-} \psi_{-}  \nn \\
& + &  \bar{\psi}_{+} \gamma^{a} \gamma_{+} \chih_{a}^{\; \mu}   i\cD_{\mu} \psi_{+} +m_3 \bar{\psi}_{+} \gamma^{3} \gamma_{+} \psi_{+} ] ,\nn \\
& - & \frac{1}{4} \chih^{\mu\mu'}\chih^{\nu \nu'} \cF_{\mu\nu}^{ab}\cF_{\mu'\nu' a'b'}  + \frac{1}{4} \bM_{\kappa} \tchi^{\mu\nu\mu'\nu'}_{aa'} \sF_{\mu\nu}^{a}\sF_{\mu'\nu'}^{a'}  \nn \\
& + &   \frac{1}{2} M_{\cA}^2 \chih^{\mu\nu} (\cA_{\mu}^{ab} -  \mOm_{\mu}^{ab}/\mu_s)( \cA_{\nu ab} -  \mOm_{\nu ab}/\mu_s) \nn \\
& + &  \frac{1}{4} \chih^{\mu\mu'}\chih^{\nu \nu'}   \cF_{\mu\nu}^{a} \cF_{\mu'\nu' a}  - \frac{1}{2} m_w^2 \chih^{\mu\nu} \cW_{\mu}^{\; a}  \cW_{\nu a}\} ,
\ee
where the spin gauge field $\cA_{\mu}^{ab}$ and chirality boost-spin gauge field $\cW_{\mu}^{\; a}$ become massive gauge bosons with mass $M_{\cA}$ and $m_w$, respectively. The covariant derivative operator involves only the spin gauge field,
\be
i\cD_{\mu} = i\p_{\mu} + \mu_{s} \cA_{\mu}^{ab} \Sigma_{ab}/2 .
\ee
This is because the gauge boson $\cW_{\mu}^{\; a}$ decouples from the qubit-spinor field due to its chirality property. 

To discuss a potential connection between the three-dimensional gravity gauge theory, as formulated within the framework of GQFT, and Witten's perspective of treating three-dimensional gravity as a gauge theory analyzed through the topological structure of Chern-Simons action, we proceed by constructing Chern-Simons action term for the GQFT in three-dimensional spacetime.

The three-dimensional Chern-Simons action is closely linked to its four-dimensional Chern class. It is useful to consider the following quadratic form for the inhomogeneous spin gauge symmetry WS(1,2):
\be
\Tr C_d \Sigma_{A\mp}\Sigma_{C\mp} & = & \Tr C_d \Sigma_{ab}\Sigma_{c\mp} + \Tr C_d \Sigma_{a\mp} \Sigma_{bc} =2 \epsilon_{abc} ,
\ee
with $C_d\equiv\gamma^3\gamma_5$ the chiral duality operation matrix in Eq.(\ref{CM}), which allows us to construct the gauge invariant Chern class as a primary characteristic class:
\be \label{GIQ}
C_4^{w} (\cA) & = & \frac {1}{2} \Tr C_d (\bscF^{-} \bscF^{-} + \bscF^{+} \bscF^{+} ) \nn \\
& = & \Tr C_d (\cF\cF^{-} + \cF\cF^{+} )  = \epsilon_{abc} \cF^{ab}\cF^{c} , 
\ee
where $\bscF^{\mp}$ is the two-form field strength defined as follows: 
\be
 & & \bscF^{\mp} \equiv d\bscA^{\mp} + \bscA^{\mp}\bscA^{\mp} \equiv \cF + \cF^{\mp} \equiv -i\mu_s\frac{1}{2} \bscF_{\mu\nu}^{\mp}dx^{\mu}\wedge dx^{\nu}, 
 \ee
with $\bscA^{\mp}\equiv -i\mu_s\bscA_{\mu}^{\mp}dx^{\mu}$ representing the one-form inhomogeneous spin gauge field. It can be verified that the above gauge invariant Chern class can be expressed into the Chern-Simons form as a secondary characteristic class: 
\be
 C_4^{w} (\cA) & = &  \epsilon_{abc} \cF^{ab}\cF^{c} =  d \omega_{3,1}^{w}(\cA, 0),
\ee
with $\omega_{3,1}^{w}(\cA, 0)$ being the Chern-Simons 3-form in three-dimensional spacetime.

The Chern-Simons action is given by the integral over the Chern-Simons 3-form: 
\be
\cS_{CS}^{w} & = & c_w\int_{M^3} \omega_{3,1}^{w}(\cA, 0) = c_w\int_{M^3}  \epsilon_{abc} \cF^{ab} \cW^{c}\nn \\
&  = &  - \frac{1}{2}\mu_s^2c_w \int d^3x \epsilon_{abc} \hat{\epsilon}^{\mu\nu\rho} \cF_{\mu\nu}^{ab} \cW_{\rho}^{\; c}   \nn \\
& = &  - \frac{1}{2}\mu_s^2c_w \int d^3x \,  \chi \, \epsilon^{a'b'c'} \chih_{a'}^{\; \mu} \chih_{b'}^{\; \nu} \chih_{c'}^{\; \rho}   \epsilon_{abc}  [ \p_{\mu}\cA_{\nu}^{ab} - \p_{\nu}\cA_{\mu}^{ab} \nn \\
& + & \mu_s(\cA_{\mu d}^{a} \cA_{\nu }^{db} - \cA_{\nu d}^{a} \cA_{\mu }^{db} ) ] \cW_{\rho}^{\; c} \equiv  \int d^3 x\, \chi \, \cL_{CS} , 
\ee
which reflects the unique property of the inhomogeneous spin gauge field in three-dimensional spacetime. In the fourth equality, we have utilized the following relation:
 \be \label{BSR}
 \hat{\epsilon}^{\mu\nu\rho}=  \chi \chih_{a'}^{\; \mu} \chih_{b'}^{\; \nu} \chih_{c'}^{\; \rho}  \epsilon^{a'b'c'} ,
 \ee
which characterizes the basic feature of three-dimensional bi-frame spacetime in GQFT, where the spin-related local orthogonal gravigauge spacetime acts as a fiber. This relation is demonstrated to be particularly useful for deriving the equation of motion for the gravigauge field.

To facilitate a direct comparison with the three-dimensional gravity treated as a gauge theory proposed in ref.\cite{GGT3D1}, it is useful to adopt the following replacements:
\be
\lambda^{a} \equiv \frac{1}{2}\epsilon^{abc}\Sigma_{bc}, \quad \cA_{\mu}^{a} \equiv \frac{1}{2}\epsilon^{a}_{\;\; bc} \cA_{\mu}^{bc}, 
\ee
with $\epsilon^{acd}\epsilon^{b}_{\;\; cd} = 2 \eta^{ab}$, and 
\be
& & [ \lambda^{a}, \lambda^{b} ] = i \epsilon^{ab}_{\;\;\;\; c} \lambda^{c}, \quad  [ \lambda_{a}, \Sigma_{b \mp} ] = - i \epsilon_{ab}^{\;\;\;\; c} \Sigma_{c\mp} , \nn \\
& &  \Tr \lambda^{a}\lambda^{b} = \eta^{ab} , \; \Tr C_d\lambda_{a}\Sigma_{b\mp} = \eta_{ab} , \;  \Tr \lambda_{a}\Sigma_{b\mp} = 0 .
\ee
With this replacement, the above Chern-Simons action can be reformulated to match the one presented in ref. \cite{GGT3D1} through the following correspondences between gauge fields:
\be
\cW_{\mu}^{\; a} \to e_{i}^{\; a}, \quad \cA_{\mu}^{a} \to \omega_{i}^{a} ,
\ee
where $e_{i}^{\; a}$ and $\omega_{i}^{a}$ are defined as the gauge fields of the inhomogeneous Lorentz gauge symmetry ISO(1,2), $A_{i} = e_{i}^{a}P_a + \omega_{i}^{a} J_a$, with $P_a$ and $J_a$ being the group generators of ISO(1,2) gauge symmetry. 

In three-dimensional gravity gauge theory proposed in ref.\cite{GGT3D1}, a pure Chern-Simons action was viewed as a unique action for gravity. Unlike in GR, the vierbein $e_{i}^{\; a}$, as a gauge field, is considered to be non-invertible. In the gravity gauge theory, both $e_{i}^{\; a}$ and $\omega_{i}^{a}$ are treated as independent gauge fields, allowing their equations of motion to describe the dynamics of three-dimensional gravity as a gauge theory. 

However, the coupling of $e_{i}^{\; a}$ to Dirac spinor field remains an open issue, as it is non-invertible and no longer transforms homogeneously under the inhomogeneous Lorentz gauge symmetry ISO(1,2). This becomes manifest from the above construction of the three-dimensional gravity gauge theory within the framework of GQFT, where the gauge fields $\cA_{\mu}^{ab}$ and $\cW_{\mu}^{\; a}$ of the inhomogeneous spin gauge symmetry WS(1,2) and the gravigauge field $\chi_{\mu}^{\; a}$ associated with the spin gauge symmetry SP(1,2) are considered as fundamental fields. Both $\cA_{\mu}^{ab}$ and $\cW_{\mu}^{\; a}$ become massive after appropriate gauge fixing conditions are applied. 

It is clear that the chirality boost-spin gauge field associated to the translation-like subgroup symmetry of WS(1,2) becomes a massive gauge field, which cannot be considered as a fundamental gravitational gauge field in GQFT. The chirality boost-spin gauge field $\cW_{\mu}^{\; a}$ can only be interpreted as a gravitational gauge field when a purely Chern-Simons action is treated as a three-dimensional gravity gauge theory, aligning with the perspective discussed in ref. \cite{GGT3D1}.

It is noteworthy that, in the absence of the Chern-Simons action, the theory exhibits a discrete $Z_2$ symmetry for the gauge field $\cW_{\mu}^{\; a}$, i.e.: 
\be
\cW_{\mu}^{\; a} \to - \cW_{\mu}^{\; a}, 
\ee
which implies that when the chirality boost-spin gauge boson acquires a mass smaller than that of the spin gauge field and qubit-spinor field, i.e., $m_w < M_{\cA}, m_3$, this gauge boson becomes a stable particle, governed by the following equation of motion:
\be
& & \cD_{\mu} \hcF^{\mu\nu }_{a}  + m_w^2 \chi \chih^{\nu\rho}\cW_{\rho a} =  \hat{\cC}_{a}^{\; \nu} , \nn \\
& & \hat{\cC}_{a}^{\; \nu}  \equiv - \frac{1}{2}\mu_s^2c_w  \hat{\epsilon}^{\nu\rho\sigma} \cF_{\rho\sigma}^{bc} \epsilon_{bca} ,
%& & \quad\; \equiv - \frac{1}{2}\mu_s^2c_w  \chi \chih_{a'}^{\; \nu} \chih_{b'}^{\; \rho} \chih_{c'}^{\; \sigma}  \epsilon^{a'b'c'}\cF_{\rho\sigma}^{bc} \epsilon_{bca}, 
\ee
with $\hcF^{\mu\nu }_{a} \equiv \chi \chih^{\mu\mu'}\chih^{\nu \nu'}\cF_{\mu'\nu' a}$. Here, the Chern-Simons action introduces a current that serves as the source for the dynamics of the chirality boost-spin gauge boson. Notably, it becomes particularly intriguing that a stable light chirality boost-spin gauge boson, as a bi-covariant vector field, may be regarded as a massive dark graviton. It has been shown that in the general theory of standard model (GSM) in four-dimensional spacetime, the massive dark graviton plays a role as a dark matter candidate\cite{GSM}.

The equation of motion for the spin gauge field is derived as follows:
\be \label{EoMSGF}
& & \cD_{\nu} \wh{\cF}_{ab}^{\mu\nu} + M_{\cA}^2 \chi \chih^{\mu\nu} (\cA_{\nu ab} - \mOm_{\nu ab}/\mu_s ) = \wh{J}_{ab}^{\mu} ,
\ee
with the field strength and current given by, 
\be
& & \wh{\cF}_{ab}^{\mu\nu}  \equiv  \chi \chih^{\mu\mu'}\chih^{\nu\nu'} \cF_{\mu'\nu' ab} , \;\;  \cD_{\nu} \wh{\cF}_{ab}^{\mu\nu} \equiv \p_{\nu} \wh{\cF}_{ab}^{\mu\nu} - \mu_s \cA_{\nu a}^{c} \wh{\cF}_{cb}^{\mu\nu}- \mu_s \cA_{\nu b}^{c} \wh{\cF}_{ac}^{\mu\nu}, \nn \\
& & \wh{J}_{ab}^{\mu} \equiv  \chi \chih^{c\mu} \mu_s\frac{1}{4} ( \bar{\psi}_{-}\Sigma_{cab}\psi_{-}  + \bar{\psi}_{+}\Sigma_{cab}\psi_{+} ) + \hat{\cC}_{ab}^{\mu} , \nn \\
& & \hat{\cC}_{ab}^{\mu}   \equiv  c_w \mu_s^2 \cD_{\nu}(\cW_{\rho}^{\; c} \hat{\epsilon}^{\mu\nu\rho} \epsilon_{abc}), 
\ee
where the Chern-Simons action contributes to the current that governs the dynamics of the spin gauge field.

Let us now to derive the gravitization equation in the presence of the Chern-Simons action. To do so, it is useful to express the Chern-Simons action in gravigauge spacetime. By recognizing the equivalence of the exterior derivative in bi-frame spacetime, $\dbar = \dbar\zeta^a\wedge \eth_{a} = d x^{\mu}\wedge\p_{\mu} = d$, we can rewrite the Chern-Simons action in gravigauge spacetime as follows:
\be
\cS_{CS}^{w} & = & - \frac{c_w}{2}\mu_s^2\int_{M^3} [d\zeta]\, \epsilon_{abc} \epsilon^{a'b'c'} \hcF_{a'b'}^{ab} \cW_{c'}^{\; c}  \nn \\
& = &  -\frac{1}{2}\mu_s^2c_w \int_{M^3} [\dbar\zeta] \epsilon_{abc} \epsilon^{a'b'c'} [ \eth_{a'}\hcA_{b'}^{ab} - \eth_{b'}\hcA_{a'}^{ab} \nn \\
& + & \mu_s(\hcA_{a' d}^{a} \hcA_{b' }^{db} - \hcA_{b' d}^{a} \hcA_{\mu }^{db} )  + \hsF_{a'b'}^{d}\hcA_{d}^{ab}] \hcW_{c'}^{\; c} . 
\ee

Following the same analyses and discussions as in the previous section on the general theory of QED, we obtain a generalized gravitization equation in the GQFT of three-dimensional spacetime,
\be \label{GE2}
\bscM_{cda}^{\; c'd'a'} \hsF_{c'd' a'}  = \widehat{\bscF}_{cda} , 
\ee
with $\bscM_{cda}^{\; c'd'a'}$ and $\widehat{\bscF}_{cda}$ given explicitly by,
\be
& & \bscM_{cda}^{\; c'd'a'} \equiv \etab_{cda}^{\; c'd'a'} \bM_{\kappa} +  \eta_{cda}^{\; c'd'a'} M_{-} + \eta_{cd}^{\; c'd'} ( \eta_{a}^{\, a'}  M_{+}   - \widehat{\cV}_{a b} \eta^{b a'} ) , \nn \\
& & \widehat{\bscF}_{cda}  \equiv   \cF_{cd}^{a'b'}\hcA_{a a'b'} +  \cF_{cd}^{a'} \hcW_{a a'}  + \frac{1}{2}\mu_s^2c_w \epsilon_{cd}^{\;\;\;\; d'}  \cW_{d'}^{\; c'} \epsilon^{c'a'b'} \cA_{a}^{a'b'}  \nn \\
& & \qquad \quad - \mu_s^{-1} M^2_{\cA} [ \hcA_{acd}  + 2(\hcA_{cda}-\hcA_{dca} ) ]  \nn \\
& & M_{\pm}  \equiv \bM_{\kappa} + (\frac{1}{2}\pm 1) M_{\cA}^2/\mu_s^2 , \;\; \widehat{\cV}_{a b}  \equiv  \hcA_{a a'b'}\hcA_{b}^{a'b'} +  \hcW_{a a'}\hcW_{b}^{\; a'} ,
\ee
which indicates that the field strength $\hsF_{cd}^{a}$ characterizing the gravitational effect in gravigauge spacetime is governed by the dynamics of the inhomogeneous spin gauge field.  

Analogues to the general theory of QED, we are able to derive the following gauge-type gravitational equation of three-dimensional GQFT from the equation of motion of the gravigauge field $\chi_{\mu}^{\; a}$ in the presence of Chern-Simons action:
\be  \label{GaGE2}
 \p_{\mu} \whsF^{\mu\nu }_{a}  = \whmJ_{a}^{\; \nu}   ,\qquad \p_{\nu} \whmJ_{a}^{\; \nu} = 0, 
\ee
with the conserved current given by,
\be \label{GC2}
\whmJ_{a}^{\; \nu}  & \equiv &16\pi G_{\kappa} \chi [  \chih_{a}^{\; \nu} \cL_{3D} -\chih_{b}^{\; \nu}  \chih_{a}^{\;\rho} \frac{1}{4} ( \bar{\psi}_{-}\gamma^{b} \gamma_{-} i\cD_{\rho} \psi_{-}   \nn \\
& + & \bar{\psi}_{+}\gamma^{b} \gamma_{+} i\cD_{\rho} \psi_{+}  ) + \chih_{a}^{\; \; \rho}  \chih^{\nu\sigma} \chih^{\rho'\sigma'} (\cF_{\rho\rho'}^{bc} \bscF_{\sigma\sigma' bc}  - \cF_{\rho\rho'}^{b} \cF_{\sigma\sigma' b} ) ]  \nn \\
 & - &  \chih_{a}^{\; \rho}  \sF_{\rho\sigma}^{b}  \whsF^{\nu\sigma}_{b} - \gamma_{W} \cD_{\rho}\widehat{\cG}_{a}^{\rho\nu} - \gamma_{W} \chih_{a}^{\; \rho}  \cG_{\rho\sigma}^{b} \widehat{\cG}^{\nu\sigma}_{b}  +  \widehat{\cC}_{a}^{\; \nu} , \nn \\
 \widehat{\cC}_{a}^{\; \nu}  & \equiv & 16\pi G_{\kappa} \chi [ \chih_{a}^{\; \nu}  \cL_{CS} + \frac{3}{2}c_w \mu_{s}^2 \chih_{a}^{\; \; \rho} \chih_{a'}^{\; \; \nu} \chih_{b'}^{\; \; \sigma}\chih_{c'}^{\; \; \lambda} \epsilon^{a'b'c'}\cF_{\rho\sigma}^{bc} \cW_{\lambda}^{\; d} \epsilon_{bcd} ],  
\ee
and the constant parameters defined by,
\be
& & 16\pi G_{\kappa} \equiv \frac{1}{\bM_{\kappa}}, \quad \gamma_W \equiv  \frac{M_{\cA}^2}{\mu_s^2\bM_{\kappa}}  .
\ee
Here, $\whsF^{\nu\sigma}_{b} \equiv \chi \1 \tchi^{[\mu\nu]\rho\sigma}_{a b}  \sF_{\rho\sigma }^{b}$ and $\widehat{\cG}_{a}^{\rho\nu}\equiv \chi\, \bchi^{[\mu\nu]\mu'\nu'}_{a a'}  \cG_{\mu'\nu' }^{a'}$ have the same definitions as shown in Eqs.(\ref{GGFS}) and (\ref{GCGFS}), respectively. The spin gauge covariant gravigauge field strength $\cG_{\mu\nu}^{a}$ is defined by, 
\be \label{SGCGFS}
& &  \cG_{\mu\nu}^{a} \equiv  \p_{\mu} \chi_{\nu}^{\; a}  - \p_{\nu} \chi_{\mu}^{\; a} + \mu_s(\cA_{\mu b}^{a} \chi_{\nu}^{\; b}  - \cA_{\nu b}^{a} \chi_{\mu}^{\; b} ).
\ee
It is evident that the Chern-Simons action gives rise to a generalized current with the additional contribution $\widehat{\cC}_{a}^{\; \nu}$. 

Meanwhile, the zero energy-momentum tensor theorem leads to the following general geometric-type gravitational equation:
 \be
 R_{\mu\nu} - \frac{1}{2} \chi_{\mu\nu} R  +  \gamma_W \cG_{\mu\nu} = 8\pi G_{\kappa} (\mT_{\mu\nu} + \mC_{\mu\nu} ),
 \ee
 where the tensors $\mT_{\mu\nu}$ and $\mC_{\mu\nu}$ are explicitly given as follows:
 \be \label{CC2}
\mT_{\mu\nu} & = &  \frac{1}{2} ( \chi_{\mu\nu} \chih_{a}^{\; \rho}   - \eta_{\mu}^{\; \rho}  \chi_{\nu a} ) ( \bar{\psi} \gamma^{a} i\cD_{\rho} \psi + H.c.) - \chi_{\mu\nu} m_3  \bar{\psi}\psi  \nn \\
& + & ( \eta_{\mu}^{\; \rho}\eta_{\nu}^{\; \sigma} - \frac{1}{4} \chi_{\mu\nu}  \chih^{\rho\sigma} )\chih^{\rho'\sigma'}  (-\cF_{\rho\rho'}^{a}\cF_{\sigma\sigma' a} + \cF_{\rho\rho'}^{ab} \cF_{\sigma\sigma' ab} ) , \nn \\ 
\mC_{\mu\nu} & \equiv & \frac{3}{2}c_w \mu_{s}^2 ( \eta_{\mu}^{\rho} \chi_{\nu a'} - \frac{1}{3} \chi_{\mu\nu} \chih_{a'}^{\; \; \rho} )   \chih_{b'}^{\; \; \sigma}\chih_{c'}^{\; \; \lambda} \epsilon^{a'b'c'}\cF_{\rho\sigma}^{bc} \cW_{\lambda}^{\; d} \epsilon_{bcd} . 
\ee
In obtaining the above general gravitational equation, the gauge-gravity-geometry relations and identities presented in Eqs. (\ref{GGGR}) and (\ref{GGGI}) are utilized. 

It is noted that the novel contribution to the source tensor, denoted by $\mC_{\mu\nu}$, originates from the Chern-Simons action. In general, the Chern-Simons action is independent of the metric, making the theory inherently topological, while it is linked to the spin-related gravigauge field $\chi_{\mu}^{\; a}$ in GQFT through the bi-frame spacetime relation provided in Eq.(\ref{BSR}).

The above general gravitational equation can be rewritten into two equations corresponding to the symmetric and antisymmetric parts:
 \be
 R_{\mu\nu} - \frac{1}{2} \chi_{\mu\nu} R  +  \gamma_W \cG_{(\mu\nu)} & = & 8\pi G_{\kappa} (\mT_{(\mu\nu)} + \mC_{(\mu\nu)} ), \nn \\
  \gamma_W \cG_{[\mu\nu]} & = & 8\pi G_{\kappa} (\mT_{[\mu\nu]} + \mC_{[\mu\nu]} ),
 \ee

Unlike GR in three-dimensional spacetime, where gravity becomes trivial due to the absence of independent degrees of freedom, the action in Eq. (\ref{3Daction4}) as a GQFT of three-dimensional spacetime is no longer trivial. It can be demonstrated, following the analyses and discussions in ref. \cite{GQFT3}, that when all source fields are ignored, the linearized gravidynamics of the gravigauge field in such a GQFT of three-dimensional spacetime involves two independent degrees of freedom. This leads to nonvanishing gravitational waves with two polarizations: scalar-like and vector-like modes, although there is no tensor-like polarization, as is well-known in GR of three-dimensional spacetime. This result arises due to the presence of the spin gauge invariant mass-like term $M_{\cA}$, indicating that the equivalence principle of GR no longer holds in GQFT. The effect is evident from the gravitational equations, which acquire a new term proportional to $\gamma_{W}$.

\section{Conclusions}

By using the general theory of QED constructed within the framework of GQFT as an example, we have demonstrated that the spin gauge symmetry of the Dirac spinor field necessitates the introduction of a spin-related gravigauge field as a bi-covariant vector field, in addition to the spin gauge field. Such a gravigauge field enables the definition of a gravigauge derivative operator and a displacement vector, which span a spin-related intrinsic gravigauge spacetime as a fiber. This forms a fiber bundle structure of bi-frame spacetime, with globally flat Minkowski spacetime serving as the base spacetime. The group structure factor of the non-commutative gravigauge derivative operator reflects the field strength of the gravigauge field in gravigauge spacetime. This field strength is shown to appear as an auxiliary field in the action of the general theory of QED formulated in local orthogonal gravigauge spacetime, allowing us to derive a constraint equation for the field strength of the gravigauge field, referred to as the gravitization equation. It indicates that the gravitational effect in such a local orthogonal gravigauge spacetime emerges from the non-commutative nature of the gravigauge derivative operator.

When the action of the general theory of QED is expressed within the framework of GQFT based on global flat Minkowski spacetime, the translational invariance across the entire Minkowski spacetime leads to a cancellation law for the energy-momentum tensor in GQFT when applying the equations of motion for all fundamental fields. This essentially extends the conservation law of the energy-momentum tensor in QFT. As a result, the total energy-momentum tensor in the entire Minkowski spacetime becomes zero due to the cancellation of contributions from the fundamental gravigauge field and other basic fields. This allows us to conclude that the gauge-type gravitational equation of the gravigauge field is equivalent to the condition of a zero energy-momentum tensor.

It has been alternatively demonstrated that the zero energy-momentum tensor theorem, as a general theorem of translational invariance within the framework of GQFT, leads to a general geometric-type gravitational equation due to the gauge-gravity-geometry correspondence. This fundamentally extends the Einstein equation of GR. Notably, in the spin gauge invariant tensor $\mT_{\mu\nu}$ associated with the Dirac spinor field and related to the spin gauge field, only the spin-related gravigauge field $\chi_{\mu}^{\; a}$ serves as the fundamental gravitational field, rather than the composite gravimetric field $\chi_{\mu\nu} = \chi_{\mu}^{\; a} \chi_{\nu}^{\; b} \eta_{ab}$. Thus, it is the gravigauge field $\chi_{\mu}^{\; a}$ that is identified as the massless graviton in GQFT. It was verified in ref. \cite{GQFT3} that such a massless graviton possesses five independent polarizations, as distinguished to the two polarizations in GR, due to the presence of the tensor term $\cG_{\mu\nu}$ in the general gravitational equation.

Several issues related to the Poincar\'e gauge theory of gravity have explicitly been examined. To explore a potential connection between GQFT and gravity gauge theories based on external symmetries, we have explicitly constructed a GQFT for three-dimensional spacetime incorporating to a Chern-Simons action. This construction begins with a simple qubit-spinor field consisting of two real components and adheres to the maximal coherence motion principle and the gauge invariance principle as guiding principles. The theory is based on the inhomogeneous spin gauge symmetry WS(1,2) and the inhomogeneous Lorentz symmetry PO(1,2), enabling for a direct comparison with three-dimensional gravity described by a purely Chern-Simons action as a gauge theory, as proposed in ref.\cite{GGT3D1}.

It is evident that the gravigauge field, as the fundamental gravitational field in GQFT, emerges as a gauge-type bi-covariant vector field. This field characterizes the bi-frame spacetime, with the spin-related gravigauge spacetime serving as a fiber. The gravigauge field behaves as a Goldstone-type boson, identified as a massless graviton. On the other hand, the chirality boost-spin gauge field, associated with the translation-like subgroup symmetry of WS(1,2), generally becomes a massive gauge field in GQFT. Nevertheless, it can only be regarded as a gravitational gauge field when considering a purely Chern-Simons action as a three-dimensional gravity gauge theory\cite{GGT3D1}.

It is interesting to probe new physics phenomena of GQFT via various experiments. The space-based gravitational wave detectors such as LISA\cite{LISA}, Taiji\cite{Taiji} and Tianqin\cite{Tianqin} are expected to probe new polarizations of gravigauge field predicted in GQFT.  Recently, it was shown in refs.\cite{GQFT3,CXW,GN} that the magnitude of coupling parameter $\gamma_W$ is constrained from the present experiments to be around $\gamma_W \simeq 10^{-2}\sim 10^{-6}$, which implies that the mass of spin gauge field should be less than the fundamental mass scale $\bM_{\kappa}\sim \sim \bM_P/\sqrt{2}$ ($\bM_P$ the reduced Planck mass) by an order of three, $M_{\cA} < 10^{-3} \bM_P$. The low bound on the mass of spin gauge field should directly be constrained from collider experiments. The most stringent bound was analyzed in ref. \cite{HTW} to be from $\mu^{+}\mu^{-}$ scattering amplitudes of all initial and final helicity configurations. As the spin gauge boson has a universal coupling to all leptons and quarks, it can be produced at hadronic colliders (such as LHC) through the gluon fusion channel. From its decay channels, such as dijets, $t\bar{t}$, dileptons, diphoton and diboson, the current experiment data at ATLAS\cite{A1,A2,A3,A4,A5} and CMS\cite{C1,C2,C3,C4,C5,C6} enable to provide strong constraints and a light spin gauge boson up to the present energy scale has already be ruled out. The near-future $e^{+}e^{-}$ colliders such as CLIC\cite{CL1}, ILC\cite{IL1}, FCC-ee\cite{FCC1} and CEPC\cite{CE1,CE2} are expected to probe further the spin gauge boson and provide more stringent constraint.

\centerline{{\bf Acknowledgement}}

This work was supported in part by the National Key Research and Development Program of China under Grant No.2020YFC2201501, the National Science Foundation of China (NSFC) under Grants No.~12147103 (special fund to the center for quanta-to-cosmos theoretical physics), No.~11821505, and the Strategic Priority Research Program of the Chinese Academy of Sciences under Grant No. XDB23030100.

\end{document}